# Rapid Co-Optimization of Processing and Circuit Design to Overcome Carbon Nanotube Variations

Gage Hills, Jie Zhang, *Member, IEEE*, Max Marcel Shulaker, Hai Wei, *Student Member, IEEE*, Chi-Shuen Lee, Arjun Balasingam, H.-S. Philip Wong, *Fellow, IEEE*, and Subhasish Mitra, *Fellow, IEEE*

*Abstract*—Carbon nanotube field-effect transistors (CNFETs) are promising candidates for building energy-efficient digital systems at highly scaled technology nodes. However, carbon nanotubes (CNTs) are inherently subject to variations that reduce circuit yield, increase susceptibility to noise, and severely degrade their anticipated energy and speed benefits. Joint exploration and optimization of CNT processing options and CNFET circuit design are required to overcome this outstanding challenge. Unfortunately, existing approaches for such exploration and optimization are computationally expensive, and mostly rely on trial-and-error-based *ad hoc* techniques. In this paper, we present a framework that quickly evaluates the impact of CNT variations on circuit delay and noise margin, and systematically explores the large space of CNT processing options to derive optimized CNT processing and CNFET circuit design guidelines. We demonstrate that our framework: 1) runs over 100× faster than existing approaches and 2) accurately identifies the most important CNT processing parameters, together with CNFET circuit design parameters (e.g., for CNFET sizing and standard cell layouts), to minimize the impact of CNT variations on CNFET circuit speed with ≤5% energy cost, while simultaneously meeting circuit-level noise margin and yield constraints.

*Index Terms*—Carbon nanotube (CNT), CNT variations, delay optimization, design-technology co-optimization.

## I. Introduction

WHILE physical scaling of silicon-based field-effect transistors has improved digital system performance for decades [10], continued device scaling is becoming increasingly challenging [2]. Carbon nanotube (CNT) field-effect transistors (CNFETs) are excellent candidates for continuing to improve both performance and energy efficiency of digital systems [13]. CNFET-based very large-scale integrated (VLSI) digital systems are projected to improve energy-delay product (EDP) by an order of magnitude versus silicon-CMOS [6], [46]. Furthermore, CNFETs provide an exciting opportunity to enable monolithic 3-D integrated circuits [47], leading to additional EDP benefits for

Manuscript received March 1, 2014; revised June 13, 2014; accepted September 22, 2014. Date of publication March 23, 2015; date of current version June 16, 2015. This work was supported in part by the National Science Foundation through the NCN-NEEDS program, contract 1227020-EEC. This paper was recommended by Associate Editor Y. Cao.
G. Hills, J. Zhang, M. M. Shulaker, H. Wei, C.-S. Lee, A. Balasingam, and H.-S. P. Wong are with the Department of Electrical Engineering, Stanford University, Stanford, CA 94305 USA (e-mail: ghills@stanford.edu).
S. Mitra is with the Department of Electrical Engineering and the Department of Computer Science, Stanford University, Stanford, CA 94305 USA.
Color versions of one or more of the figures in this paper are available online at http://ieeexplore.ieee.org.
Digital Object Identifier 10.1109/TCAD.2015.2415492

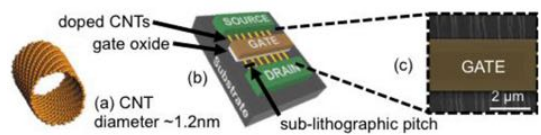

Fig. 1. (a) CNT. (b) Typical CNFET structure. (c) Scanning electron microscopy image of the CNFET channel.

CNFET-based digital systems with massive integration of logic and memory [42].

The schematic of a CNFET is shown in Fig. 1. Multiple CNTs compose the transistor channel, whose conductance is modulated by the gate. The gate, source, and drain are defined using traditional photolithography, while the CNT-CNT spacing is determined by the CNT growth [31] and can therefore exceed the minimum lithographic pitch. For high drive current, the target CNT-CNT spacing is 4–5 nm [46].

Despite demonstrations of sub-10 nm channel length CNFETs [13] and stand-alone CNFET circuit elements [5], [7], [11], realization of complex CNFET-based digital systems had been prohibited by substantial imperfections inherent to CNTs: mis-positioned CNTs and metallic CNTs. Mis-positioned CNTs cause stray conducting paths that can lead to incorrect logic functionality, and metallic CNTs (resulting from the imprecise control over CNT properties) result in increased leakage current and can lead to incorrect logic functionality. A unique combination of CNT processing and CNFET circuit design techniques, known as the imperfection-immune paradigm [54], overcomes these challenges in a VLSI-compatible manner to enable the realization of the first CNFET-based digital systems [32], [33], [40], including the first programmable microprocessor built using CNFETs [39]. Two key enablers of these demonstrations are: 1) mis-positioned CNT-immune layout design [30] and 2) VLSI-compatible metallic CNT removal (VMR), which efficiently removes ≥99.99% of metallic CNTs [32], [40].

Unfortunately, process variations specific to CNTs, such as the imprecise control over CNT properties and the nonuniform density of grown CNTs (details in Section II), can lead to significantly reduced circuit yield, increased susceptibility to noise, and large variations in CNFET circuit delays (Section II) [54]. One method to counteract these effects is to upsize all CNFETs. However, such naïve upsizing incurs large energy and delay costs that diminish CNFET technology benefits.

Rather, various CNT process improvement options, when combined with CNFET circuit design, provide an energy-efficient method of overcoming CNT variations. Without such strategies, CNT variations can degrade the potential speed benefits of CNFET circuits by $\geq 20\%$ at sub-10 nm nodes, even for circuits with upsized CNFETs to achieve $\geq 99.9\%$ yield (Section II). By leveraging CNT process improvements, together with CNFET circuit design, the overall speed degradation can be limited to $\leq 5\%$ with $\leq 5\%$ energy cost while simultaneously meeting circuit-level noise margin and yield constraints [52].

However, co-optimization of CNT technology options and CNFET circuit design parameters using trial-and-error-based search can be prohibitively time-consuming. In this paper, we demonstrate a systematic and VLSI-scalable methodology that selects effective combinations of CNT processing options and CNFET circuit design techniques to overcome CNT variations. Our key contributions are as follows.

1) Techniques to quickly evaluate the impact of CNT variations on circuit yield, susceptibility to noise, delay, and energy. They run $>100\times$ faster than previous approaches.
2) A systematic methodology to explore the large space of CNT processing options together with CNFET circuit design parameters (e.g., CNFET sizing and standard cell layouts leveraging CNT correlation, see Section II), to rapidly identify designs that reduce the impact of CNT variations on circuit yield, susceptibility to noise, and delay variations with $\leq 5\%$ energy cost. This is in sharp contrast to previous trial-and-error-based approaches.
3) Derivation of guidelines for CNT processing and CNFET circuit design parameters at highly scaled technology nodes to overcome CNT variations. We provide guidelines to limit the overall circuit speed degradation to $\leq 5\%$ with $\leq 5\%$ energy cost while maintaining $\geq 99.999\%$ functional circuit yield and $\leq 0.001\%$ probability of failing to meet circuit-level noise margin requirements (Section IV).

In Section II, we present an overview of CNT variations and their impact on CNFET circuits. Section III describes a methodology to optimize circuit performance in the presence of CNT variations, leveraging a SPICE-compatible CNFET device model to build efficient variation-aware models for the delay, energy, and noise margin of CNFET circuits. Using this methodology, we provide CNT processing and CNFET circuit design guidelines for overcoming CNT variations at the 14, 10, 7, and 5 nm technology nodes (Section IV).

An earlier version of this paper was published in [16]. Here, we present the following additional contributions.
1) Design and analysis of CNFET digital VLSI circuits scaled to the 5 nm node, enabled by a recently developed SPICE-compatible CNFET device model for accurate analysis of sub-10 nm gate length CNFETs [23].
2) A computationally efficient technique to numerically calculate the probability that CNFET circuits fail to meet circuit-level noise margin requirements. This technique can accurately compute such probabilities less than 0.001% (as is desirable for VLSI-scale circuits, details in Section II-C).

In this paper, we make references to [17], which contains additional figures and analysis details. It is available for download at http://www.arxiv.org.

TABLE I
CNT PROCESSING PARAMETERS FOR CNT COUNT VARIATIONS. CNT DENSITY = 250 CNTs/$\mu$M FOR ALL ANALYSIS [53]

| Proc. Param | Definition | Ideal value | Experimental value |
|---|---|---|---|
| $IDC$ | Index of Dispersion for CNT count $IDC = \sigma_s^2 / \mu_s$ $\mu_s$ and $\sigma_s^2$: mean and variance of the distribution of CNT-CNT spacing [49] | 0 | 0.50 [49] |
| $p_m$ | Probability that a given CNT is an m-CNT | 0% | 1%-10% [28], [29] |
| $p_{Rs}$ | Conditional probability that a CNT is removed, given that it is an s-CNT | 0% | 4% [40] |
| $p_{Rm}$ | Conditional probability that a CNT is removed, given that it is an m-CNT | 100% | > 99.99% [40] |

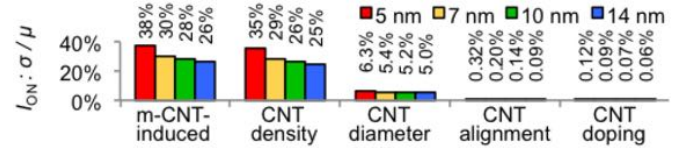

Fig. 2. CNFET $I_{ON}$ variations due to CNT variations (x-axis) for a minimum-width CNFET ($V_{DD} = 0.50$ V, width = half-contacted gate pitch; see [17, Table VI]). $IDC = 0.50$ [49] for CNT density variations, $p_m = 33\%$ [37] and $p_{Rs} = 4\%$ [40] for m-CNT-induced variations. Diameter is normally distributed with $\mu_d = 1.3$ nm and $\sigma_d = 0.1$ nm [31]. Alignment and doping distribution details in [54]. To analyze $I_{ON}$ variations attributed to individual sources of CNT variations, all other sources of CNT variations are removed. Additional parameters in [17, Table VI].

## II. CNT VARIATIONS

In addition to process variations that exist for silicon-CMOS FETs (e.g., variations in channel length, oxide thickness, and threshold voltage [26]), CNFETs are also subject to CNT-specific variations, including variations in CNT type (semiconducting: s-CNT or metallic: m-CNT) [32], CNT density [49], diameter [34], alignment [30], and doping [9] (details in [17, Sec. VI]). While the on-current ($I_{ON}$) of a CNFET with only a single CNT as its channel is highly sensitive to CNT diameter variations [34], CNFETs in practical VLSI circuits consist of multiple CNTs to provide sufficient $I_{ON}$. Thus, the impact of diameter variations is reduced due to statistical averaging (Fig. 2) [35]. Rather, $I_{ON}$ variations are dominated by variations in the CNT count: the number of s-CNTs in a CNFET (after m-CNT removal, e.g., using VMR)[1] [52]. CNT count variations stem from two sources.

1) *CNT Density Variations:* Precise positioning of CNTs is difficult to control; resulting CNT-CNT spacing variations lead to a variable number of CNTs in each CNFET [49].
2) *m-CNT-Induced Variations:* Each CNFET contains a variable number of both s-CNTs and m-CNTs, resulting in CNT count variations even assuming a perfectly selective m-CNT removal technique (i.e., $p_{Rm} = 100\%$, $p_{Rs} = 0\%$: Table I). In addition, m-CNT removal techniques may inadvertently remove a small fraction of s-CNTs, further contributing to CNT count variations [54].

CNT count variations are parameterized by the parameters: Index of Dispersion for CNT count ($IDC$), $p_m$, $p_{Rs}$, and $p_{Rm}$ (i.e., the *processing parameters*) defined in Table I. We analyze the impact of CNT count variations on CNFET circuit modules synthesized from the processor core of OpenSPARC T2, a large multicore chip that closely resembles the commercial Oracle/SUN Niagara 2 system [27]. These OpenSPARC modules consist of ∼4 K to >100 K logic gates (Table III) and expose several effects in VLSI-scale circuits (e.g., wire parasitics) that are not visible in small circuit benchmarks. We consider the effects of CNT count variations on the following circuit-level metrics.

---
[1]Another technique for post-growth m-CNT removal is known as CNT sorting, in which s-CNTs are separated from m-CNTs in a solution [1]. However, CNT sorting techniques have not yet achieved the selectivity required for VLSI-scale digital circuits [50].

1) *Functional Yield:* Due to CNT count variations, there is nonzero probability that a CNFET contains no s-CNTs in its channel, leading to functional failure of the CNFET (i.e., CNT count failure) [51]. The count-limited yield of a CNFET circuit is the probability that no CNFET experiences CNT count failure [51] (Section II-A).
2) *Delay Penalty:* The increase in the 95-percentile-delay ($T_{95}$: the minimum clock period that the circuit has a 95% probability of meeting) relative to the nominal delay (the critical path delay when there are no variations). Details in Section II-B.
3) *Static Noise Margin (SNM):* A measure of the noise susceptibility of a pair of connected logic gates (Section II-C).
4) *Probability of Noise Margin Violation (PNMV):* The probability that any pair of connected logic gates in a circuit fails to meet $SNM_R$, a required SNM level (Section II-C).

### A. Impact on Circuit Functional Yield

For VLSI CNFET circuits with minimum-width CNFETs, the count-limited yield can be very low (near zero) [51]. An effective method to significantly improve the count-limited yield (≥99.999%) is to perform minimum-width upsizing: upsize all CNFETs that have width ($W$) less than a specified minimum width ($W_{MIN}$) to have $W = W_{MIN}$ [51]. Although minimum-width upsizing effectively improves count-limited yield, it can incur large energy costs if the CNT count failures of all CNFETs are independent [51]. Rather, for CNFET circuits with highly aligned CNTs, the count-limited yield (and the energy cost of minimum-width upsizing, details below) can be significantly improved by leveraging the unique property of CNT correlation: since CNTs are 1-D nanostructures with lengths typically much longer than the CNFET contacted gate-pitch [20], [31], the CNT counts of CNFETs can be uncorrelated or highly correlated depending on the relative physical placement of the CNFET active regions (active region: area of channel which has CNTs) [51]. Special aligned-active layouts can engineer these correlations by aligning the active regions in a library to maximize correlation [17, Fig. 15]. Aligned-active layouts incur minimal area increase (only 4 of 134 cells from the Nangate 45 nm Open Cell Library [25] incur area penalties <14%), and the locations of I/O pins are mostly retained, resulting in negligible impact on intercell routing [51].

To achieve count-limited yield ≥99% for circuits today (which can consist of 100M logic gates), the count-limited yield for each OpenSPARC module (∼100K logic gates) should be ≥99.999%. To reach this target, we use a combination of minimum-width upsizing, aligned-active layouts, and CNT process improvements. We first use minimum-width upsizing with aligned-active layouts to achieve count-limited yield ≥99.9% (which is lower than the 99.999% requirement, details below). Then, CNT process improvements (which are required to meet delay penalty and noise margin requirements) further improve the count-limited yield, e.g., to ≥99.999% (details below in steps 1–3). We define $\Delta E_{FUNC}$ as the energy cost (in terms of total energy per cycle) of minimum-width upsizing to reach a desired count-limited yield (i.e., functional yield). $\Delta E_{FUNC}$ can be ≤2.5% for all the OpenSPARC modules (Fig. 4). It is determined using the design flow in Fig. 3. Steps 1–3 (Fig. 3) are described below.

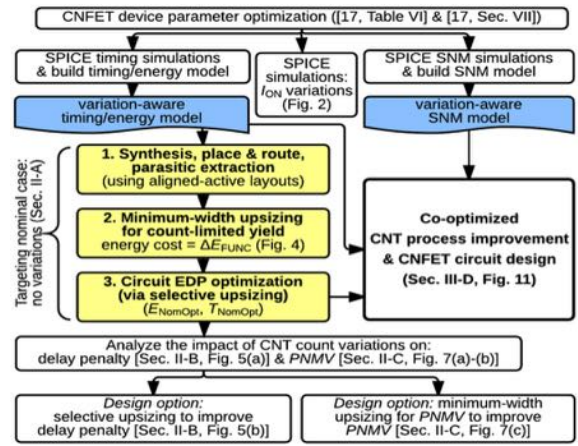

Fig. 3. Full analysis & design methodology. Steps 1–3 (highlighted) are described in this section. Additional details in Section III and [17, Sec. VII].

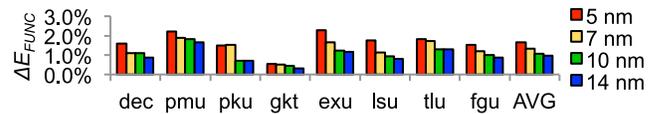

Fig. 4. Energy cost of minimum-width upsizing with aligned-active layouts to achieve ≥99.9% count-limited yield: OpenSPARC modules, $I_{DC} = 0.50$, $p_m = 10\%$, $p_{Rs} = 4\%$, $p_{Rm} = 99.99\%$ (count-limited yield improves to ≥99.999% with the processing guidelines in Section IV). Improving delay penalty and PNMV can require additional energy costs.

1) *Synthesis, Place and Route, and Parasitic Extraction:* Targeting the nominal case: no variations. Details in [17, Sec. VII].
2) *Minimum-Width Upsizing for Count-Limited Yield:* Determine $W_{MIN}$ to achieve count-limited yield ≥99.9% with aligned-active layouts via the methodology in [51], using experimentally demonstrated values for the processing parameters (Table I, though other values may be chosen). Then perform minimum-width upsizing (the associated energy cost is $\Delta E_{FUNC}$). Note that, this initial count-limited yield target of ≥99.9% is lower than the required ≥99.999% count-limited yield. In Section IV, we show that CNT process improvements (which are required to meet delay penalty and noise margin requirements) further improve the count-limited yield, e.g., to ≥99.999%. If count-limited yield ≥99.999% is not achieved after meeting delay penalty and noise margin requirements, we return to this step and increase $W_{MIN}$ to the width of the next-largest CNFET in our standard cell library (details in [17, Sec. VII]).
3) *Circuit EDP Optimization:* We use the EDP metric to quantify energy efficiency. We perform circuit sizing to minimize circuit EDP using a selective transistor/logic gate upsizing algorithm (i.e., selective upsizing) inspired by [52]: targeting the nominal case, we first sort all standard cells according to their fan-out (fan-out: the ratio of the output load capacitance to the minimum input capacitance on any input), then upsize a parameterized number $k_{SelUpsize} \geq 0$ of the standard cells with the largest fan-out (see algorithm in [17, Sec. VII-C]). We sweep $k_{SelUpsize}$ to generate an energy-delay trade-off curve. We record the nominal energy ($E_{Nom}$) and the nominal critical path delay ($T_{Nom}$) for each point on this curve, and then select the point with the minimum EDP [$EDP_{NomOpt}$, defined in (1)]. This point (for the nominal case) is referred to as the EDP-optimized nominal design point: ($E_{NomOpt}$, $T_{NomOpt}$). All delay penalties are relative to this point

$$EDP_{NomOpt} = E_{NomOpt} T_{NomOpt}. \quad (1)$$

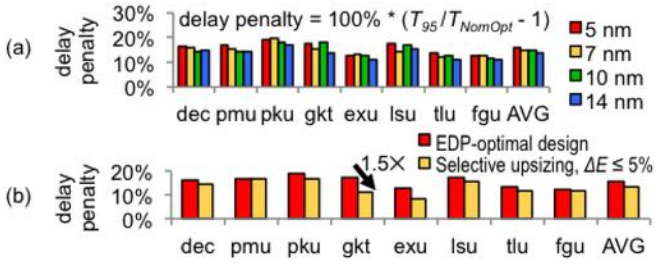

Fig. 5. Delay penalty for the OpenSPARC modules (after steps 1–3 in Fig. 3). $IDC = 0.50$, $p_m = 10\%$, $p_{Rs} = 4\%$, $p_{Rm} = 99.99\%$. (a) Delay penalty across technology nodes. (b) Delay penalty improvement due to selective upsizing with $\Delta E \leq 5\%$. For both (a) and (b): count-limited yield $\geq 99.98\%$ in all cases (it improves to $\geq 99.999\%$ with the processing guidelines in Section IV).

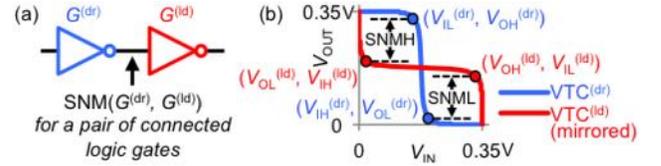

Fig. 6. SNM illustration. (a) Example gate pair. (b) $VTC^{(dr)}$ and mirrored $VTC^{(ld)}$. For each of $G^{(dr)}$ and $G^{(ld)}$: $V_{OH}$, $V_{IH}$, $V_{IL}$, and $V_{OL}$ are taken from the points on the VTC where the slope is $-1$.

While ($E_{NomOpt}$, $T_{NomOpt}$) represents an attractive design in the nominal case (since $EDP_{NomOpt}$ is small versus other points on the energy-delay tradeoff curve), this design may have a high delay penalty due to CNT variations (e.g., it can be $\geq 20\%$ at sub-10 nm nodes: Section II-B).

### B. Impact on Circuit Delay Variations

To derive distributions of CNFET circuit delays resulting from CNT count variations, we leverage the methodology described in [52]. This is a Monte Carlo statistical static timing analysis (MC SSTA) approach with two key changes: 1) a variation-aware timing model for CNFET logic gates (built using a CNFET device model [45]) and 2) highly efficient CNT count sampling, based on the unique asymmetric CNT correlation property (Section II-A). This allows us to compute the delay penalty for each OpenSPARC module (after steps 1–3 in Fig. 3) as follows: sample the delay distribution via MC SSTA (using 2000 trials, excluding any trials that have CNT count failure), then extract $T_{95}$ from the delay distribution to calculate the delay penalty [Fig. 5(a)]. Fig. 5(a) illustrates that the delay penalty for the OpenSPARC modules can be $\geq 20\%$ for EDP-optimized designs with aligned-active layouts at highly scaled technology nodes.

To overcome CNT variations, we target delay penalty $\leq 5\%$ with total energy per cycle cost $\Delta E \leq 5\%$ [relative to $E_{NomOpt}(1)$] to maintain $\geq 90\%$ of the projected EDP benefits of CNFET circuits, even in the presence of CNT variations. To improve delay penalties we leverage the selective upsizing approach described in Section II-A [52]. Fig. 5(b) shows that selective upsizing can reduce delay penalties by $1.5\times$ (e.g., from 17% to 11% for the "gkt" OpenSPARC module); in Fig. 5(b), additional selective upsizing was performed after steps 1–3 in Fig. 3 by increasing $k_{SelUpsize}$ to minimize the delay penalty subject to $\Delta E \leq 5\%$.

### C. Impact on Circuit PNMV

A common metric to quantify the noise susceptibility of a pair of connected logic gates [i.e., a gate pair: ($G^{(dr)}$, $G^{(ld)}$), where $G^{(dr)}$ and $G^{(ld)}$ are the driving and loading logic gates, respectively] is the SNM, which can be quantified as follows [using the gate pair shown in Fig. 6(a) as an example]. Let $G^{(dr)}$ have voltage transfer curve $VTC^{(dr)}$ (voltage transfer curve: $V_{OUT}$ versus $V_{IN}$ in the static case) and let $G^{(ld)}$ have voltage transfer curve $VTC^{(ld)}$. Also, let ($V_{IL}^{(dr)}$, $V_{OH}^{(dr)}$) and ($V_{IH}^{(dr)}$, $V_{OL}^{(dr)}$) be the points on $VTC^{(dr)}$ where the slope of $V_{OUT}$ versus $V_{IN}$ is $-1$ [as shown in Fig. 6(b)]. Similarly define ($V_{IL}^{(ld)}$, $V_{OH}^{(ld)}$) and ($V_{IH}^{(ld)}$, $V_{OL}^{(ld)}$) for $VTC^{(ld)}$ [mirrored in Fig. 6(b)]. Then for the gate pair ($G^{(dr)}$, $G^{(ld)}$), the high SNM (SNMH), the low SNM (SNML), and the SNM are defined in (2)–(4), respectively [48]

$$SNMH\left(G^{(dr)}, G^{(ld)}\right) = V_{OH}^{(dr)} - V_{IH}^{(ld)} \quad (2)$$

$$SNML\left(G^{(dr)}, G^{(ld)}\right) = V_{IL}^{(ld)} - V_{OL}^{(dr)} \quad (3)$$

$$SNM\left(G^{(dr)}, G^{(ld)}\right) = \min\left(V_{OH}^{(dr)} - V_{IH}^{(ld)}, V_{IL}^{(ld)} - V_{OL}^{(dr)}\right). \quad (4)$$

$SNM(G^{(dr)}, G^{(ld)})$ is sensitive to $I_{ON}$ variations [48], and so it is sensitive to CNT count variations. To quantify the impact of SNM variations on circuit noise susceptibility, we use the *PNMV* metric, which is the probability that any gate pair in a circuit fails to meet a required SNM level: $SNM_R$. $SNM_R$ is a design constraint chosen by the designer and *PNMV* is directly related to $SNM_R$. As $SNM_R$ increases (tighter SNM requirement) then *PNMV* increases (lower probability of meeting the SNM requirement). Typical values of $SNM_R$ are relative to the supply voltage, $V_{DD}$ (e.g., $SNM_R = V_{DD}/5$ [48]). *PNMV* is defined in (5), where $C$ is the set of all gate pairs

$$PNMV = 1 - P\left\{\bigcap_{(G^{(dr)}, G^{(ld)}) \in C} \left(SNM(G^{(dr)}, G^{(ld)}) \geq SNM_R\right)\right\}. \quad (5)$$

To solve for *PNMV* due to CNT count variations, we leverage a variation-aware SNM model that can compute $SNMH(G^{(dr)}, G^{(ld)})$ and $SNML(G^{(dr)}, G^{(ld)})$ for every gate pair in a circuit, given the CNT counts of each CNFET contained in $G^{(dr)}$ and $G^{(ld)}$ (details in Section III-B1). In Section III-B2, we describe how to combine this variation-aware SNM model and the distributions of CNT count for all CNFETs in the circuit to efficiently calculate *PNMV*.

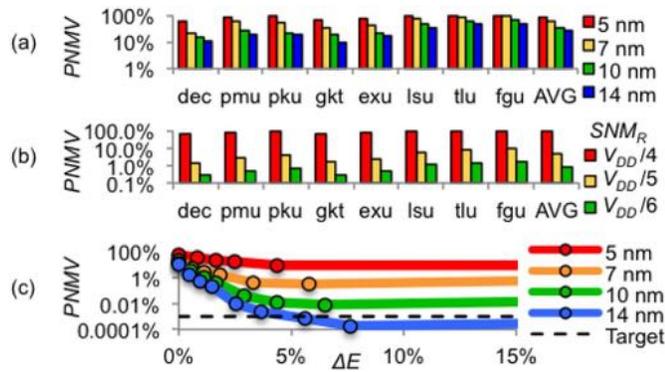

Fig. 7. *PNMV* for the OpenSPARC modules (after steps 1–3 in Fig. 3). $IDC = 0.50$, $p_m = 10\%$, $p_{Rs} = 4\%$, $p_{Rm} = 99.99\%$. (a) *PNMV* versus node ($SNM_R = V_{DD}/4$). (b) *PNMV* versus $SNM_R$ (5 nm node). (c) *PNMV* versus $\Delta E$ for additional minimum-width upsizing (in addition to minimum-width upsizing for count-limited yield, Fig. 3): "dec" OpenSPARC module. For (a), (b) and (c): count-limited yield ≥99.98% in all cases (it improves to ≥99.999% with the processing guidelines in Section IV).

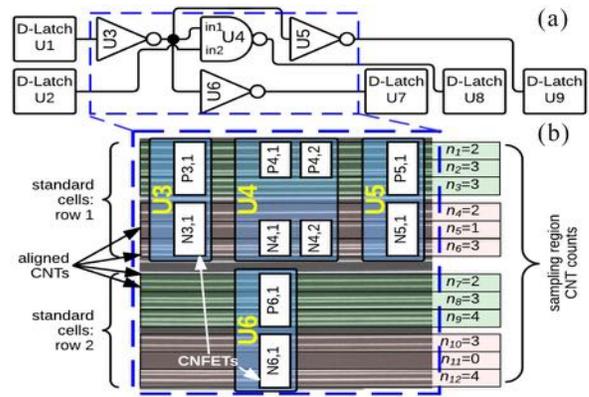

Fig. 8. (a) Subset of logic gates in an example circuit module. (b) Illustration of two rows of standard cells that depicts the relationship between the sampling region CNT counts (e.g., $n_1, n_2, \ldots, n_{12}$) and the CNT counts of each CNFET [53]. For example, the CNT count for CNFET $P3,1$ in inverter $U3$ is $n_1 + n_2 + n_3 = 2 + 3 + 3 = 8$.

Fig. 7(a) and (b) quantify *PNMV* for the OpenSPARC modules (after steps 1–3 in Fig. 3), which can be nearly 100% at the 5 nm node. To achieve *PNMV* ≤ 1% for circuits today (with ∼100M logic gates), each OpenSPARC module (∼100K logic gates) should have *PNMV* ≤ 0.001%.

Since minimum-width CNFETs are highly sensitive to CNT count variations [52], gate pairs that contain minimum width CNFETs are highly likely to cause SNM violations. Thus, *PNMV* is highly sensitive to minimum-width CNFETs, so further minimum-width upsizing (in addition to minimum-width upsizing for count-limited yield) improves *PNMV* (via statistical averaging) at the cost of energy [Fig. 7(c)]. However, additional minimum-width upsizing may be undesirable as it can require $\Delta E > 5\%$, can increase circuit delay, and is not guaranteed to meet *PNMV* constraints [17, Sec. IX-A].

### D. Overcoming CNT Variations

As shown above, CNFET upsizing techniques alone can be insufficient to meet design goals (e.g., delay penalty ≤5% and *PNMV* ≤ 0.001% with $\Delta E \leq 5\%$) [54]. Rather, a combination of CNT processing and CNFET circuit design is required [54], but two key questions must be answered: 1) which processing parameters to improve? 2) By how much?

Without a systematic methodology to evaluate the circuit-level impact of CNT variations, one might blindly pursue difficult CNT processing paths with diminishing returns, while overlooking other processing parameters that enable larger performance gains. For example, much research has focused solely on improving $p_m$ [1]. However, reducing $p_m$ past 1% suffers from diminishing returns and can be insufficient to meet design goals [16], [54] (e.g., in Fig. 16 in [17, Sec. VII]: $p_m = 0.1\%$ does not achieve delay penalty ≤5%).

Previously, co-optimization of processing and design has been performed via a trial-and-error-based approach [52]. However, this can be prohibitively time-consuming, potentially requiring months of simulation time (details in Section IV). In Section III, we present a methodology that efficiently selects effective combinations of CNT processing options and CNFET circuit design techniques to overcome CNT variations.

### III. RAPID CO-OPTIMIZATION OF PROCESSING & DESIGN

An existing approach to overcome CNT variations is based on brute-force trial-and-error [52]: a designer iterates over many design points (design point: a combination of values for the CNT processing parameters: *IDC*, $p_m$, $p_{Rs}$, $p_{Rm}$, and the CNFET design parameters: e.g., $k_{SelUpsize}$), analyzing each one until a design point that satisfies a target delay penalty and target *PNMV* with small energy cost is found. Furthermore, this approach utilizes highly accurate yet computationally expensive models to calculate delay penalties and *PNMV*. It suffers from two significant bottlenecks.
1) The time required to calculate delay penalties and *PNMV* limits the number of design points that can be explored.
2) The number of required simulations can be exponential in the number of CNT processing and CNFET design parameters.

Our methodology overcomes these bottlenecks as follows.
1) We estimate delay penalties >100× faster than the previous approach and efficiently calculate *PNMV* ≤ 0.001%, enabling exploration of many more design points while maintaining sufficient accuracy to make correct design decisions (details in Section IV).
2) We use a gradient descent search algorithm, based on delay and *PNMV* sensitivity information with respect to the processing parameters, to systematically guide the exploration of design points (details in Section III-D).

### A. Rapid Quantification of Circuit Delay Penalty

To quantify CNFET circuit delay variations, we leverage the probabilistic framework in [52], which is based on an MC SSTA approach with two key enhancements.
1) *Highly Efficient Sampling Method:* It is not trivial to analytically model the effects of CNT correlation at the circuit level. We partition the circuit area in sampling regions, each of which has its own independent CNT count. The CNT count of each CNFET is then the sum of the CNT counts of each sampling region that it overlaps (example shown in Fig. 8) [53].
2) *Variation-Aware Timing Model:* The drive current and parasitic capacitances of CNFETs are modeled as affine functions of the CNT counts in each sampling region [53].

TABLE II
VARIABLES IN THE FULL-CIRCUIT DELAY MODEL

| Variable | $\in$ | Equation | Description |
|---|---|---|---|
| $Q_{MC}$ | $\mathbb{R}^{m \times n}$ | $V_{DD}A_{CLoad}X$ | Variable charge per CNT |
| $q_{Exp}$ | $\mathbb{R}^m$ | $V_{DD}A_{CLoad}\mathbf{1}$ | Expected charge per CNT |
| $q_{Fix}$ | $\mathbb{R}^m$ | $V_{DD}b_{CLoad}$ | Fixed charge |
| $I_{MC}$ | $\mathbb{R}^{m \times n}$ | $A_{IDrive}X$ | Variable current per CNT |
| $i_{Exp}$ | $\mathbb{R}^m$ | $A_{IDrive}\mathbf{1}$ | Expected current per CNT |
| $i_{Fix}$ | $\mathbb{R}^m$ | $b_{IDrive}$ | Fixed current |

We incorporate two additional enhancements to improve computation time and for sensitivity analysis of CNFET circuit delay variations versus each processing parameter.

1) *Gaussian Approximation of CNT Count Distributions:* This allows us to factor the variation-aware timing model into two components, one of which does not depend on the processing parameter values and can therefore be precomputed (details below). The CNT count variables are thus elements of the set of real numbers ($\mathbb{R}$) instead of the set of nonnegative integers ($\mathbb{Z}^+$) [49]. The accuracy of this approximation is validated in [17, Sec. VI-B].

2) *Linearized Timing Model for Delay Variations:* We leverage the same timing model as in [53] to compute the maximum path delay of a circuit when no variations are present (nominal delay). Then, we linearize this nonlinear timing model (around the nominal case), and use the resulting linearized timing model to analyze the impact of CNT variations on CNFET circuit delay variations (details in [17, Sec. VIII]). Similar techniques are often used to approximate silicon-CMOS-based circuit delays in early design stages [24]. Unlike in [53], we fix the input slew rate of each logic gate to its nominal value. This allows us to efficiently compute all of the logic gate delays in a circuit simultaneously. These approximations have minimal impact on our design choices (Section IV). We refer to the model in [53] as the nonlinear timing model, and to the model described below as the linearized timing model.

To formulate the delay model for the full circuit, let $\mu_R$ and $\sigma_R$ be the mean and standard deviation of the sampling region CNT count distribution ($\mu_R$ and $\sigma_R$ are functions of the processing parameters shown in Table I). The first step to estimate the delay penalty of a design point is to sample the CNT count for each sampling region and for each MC trial. Each sample is one entry in a matrix $N \in \mathbb{R}^{r \times n}$, where $r$ is the total number of sampling regions and $n$ is the total number of MC trials. We then compute the total capacitive load and drive current for each of the $m$ gates (for each trial) via an affine transformation of the region CNT counts (based on the model in [53]). We express this transformation in matrix form, where $C_{Tot}, I_{Drive} \in \mathbb{R}^{m \times n}$

$$C_{Tot} = A_{CLoad}N + b_{CLoad}\mathbf{1}^T \qquad (6)$$
$$I_{Drive} = A_{IDrive}N + b_{IDrive}\mathbf{1}^T. \qquad (7)$$

Our delay models are fully specified by $A_{CLoad}, A_{IDrive} \in \mathbb{R}^{m \times r}$ and column vectors $b_{CLoad}, b_{IDrive} \in \mathbb{R}^m$, which contain the coefficients of the affine transformations from the sampling region CNT counts to the CNFET drive currents and parasitic capacitances [53]. Next, we factor out $\mu_R$ and $\sigma_R$, a crucial step in achieving computational efficiency. We rewrite $N = \mu_R \mathbf{1}\mathbf{1}^T + \sigma_R X$, where each element of $X \in \mathbb{R}^{r \times n}$ is distributed according to a unit Gaussian distribution, allowing (6)-(7) to be written as

$$C_{Tot} = \sigma_R A_{CLoad}X + (\mu_R A_{CLoad}\mathbf{1} + b_{CLoad})\mathbf{1}^T \qquad (8)$$
$$I_{Drive} = \sigma_R A_{IDrive}X + (\mu_R A_{IDrive}\mathbf{1} + b_{IDrive})\mathbf{1}^T. \qquad (9)$$

Note that, $\mathbf{1}$ is a column vector with every element equal to 1, and multiplication of a matrix by a scalar (e.g., $\mu_R$ or $\sigma_R$) indicates that each element in the matrix is multiplied by that scalar. Any product that does not contain $\mu_R$ or $\sigma_R$ is independent of the processing parameters, and can therefore be precomputed. The dominant computational tasks are the matrix multiplications $AX$ {which are $O(mn)$ since $A$ is sparse [12]}. Precomputing such terms (and factoring in the multiplication of $C_{Tot}$ and $V_{DD}$), yields equivalent expressions for total charge and drive currents that require scalar operations (see Table II for variable definitions)

$$Q_{Tot} = \sigma_R Q_{MC} + (\mu_R q_{Exp} + q_{Fix})\mathbf{1}^T \qquad (10)$$
$$I_{Drive} = \sigma_R I_{MC} + (\mu_R i_{Exp} + i_{Fix})\mathbf{1}^T. \qquad (11)$$

Precomputing $Q_{MC}$, $q_{Exp}$, $q_{Fix}$, $I_{MC}$, $i_{Exp}$, and $i_{Fix}$ (Table II) subsequently allows each logic gate delay to be efficiently computed with only two multiplications, one division, and three additions per trial [only counting operations in (12) that must be computed for each trial]. This includes the addition of $d_{Fix} \in \mathbb{R}^m$, a vector of fixed delays (e.g., input delays from external circuits). The matrix division in (12) is element-wise

$$D = \frac{\sigma_R Q_{MC} + (\mu_R q_{Exp} + q_{Fix})\mathbf{1}^T}{\sigma_R I_{MC} + (\mu_R i_{Exp} + i_{Fix})\mathbf{1}^T} + d_{Fix}\mathbf{1}^T. \qquad (12)$$

We then perform static timing analysis (STA) for each MC trial (and for the nominal case), and use the results to estimate $T_{95}$ and the delay penalty. The total circuit energy is computed using a model of the form $E = (1/2)CV^2$ [48]

$$E_{Tot} = (1/2)V_{DD}\mathbf{1}^T((1/n)\sigma_R Q_{MC}\mathbf{1} + \mu_R q_{Exp} + q_{Fix}). \qquad (13)$$

### B. Rapid Quantification of Circuit PNMV

Our method of analyzing circuit *PNMV* consists of two key components, each of which is described in this section.

1) A variation-aware SNM model, which computes $V_{OH}$, $V_{IH}$, $V_{IL}$, and $V_{OL}$ (these terms are defined in Section II-C) as functions of the CNT counts of the CNFETs within a logic gate. This model can be used to compute SNM for every gate pair in the circuit.

2) A method to numerically calculate low *PNMV* values (e.g., ≤0.001%), given the variation-aware SNM model and given a network of cascaded logic gates (e.g., a circuit module after steps 1–3 in Fig. 3). This technique accounts for correlations in CNT count among CNFETs.

*1) Variation-Aware Static Noise Margin Model:* We refer to $V_{OH}$, $V_{IH}$, $V_{IL}$, and $V_{OL}$ as the VTC parameters, and we model them for each stage of cascaded logic. We distinguish logic stages from standard cells since a standard cell can consist of multiple logic stages (e.g., the standard cell BUF_X1 consists of two cascaded inverters, each of which is one logic stage). For standard cells with multiple logic stages, we model the VTC parameters separately for each logic stage (e.g., we consider the cross-coupled inverters in a D-latch as two separate logic stages). For consistency with the terminology in

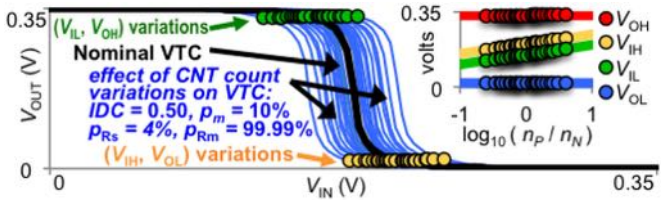

Fig. 9. Variations in the VTC due to CNT count variations (shown for an inverter at the 5 nm node with $V_{DD} = 0.35$ V: e.g., inverter $U3$ in Fig. 8, with $n_P = n_P^{(P3,1)}$ and $n_N = n_N^{(N3,1)}$). Example VTCs: simulated using SPICE by sweeping $V_{IN}$ to obtain $V_{OUT}$. (Inset) VTC parameters versus CNT count. Markers represent extracted values of the VTC parameters, lines represent the SNM model (14) for $T^{(INV\_X1)}$ with $T_{VOH0} = 0.33$, $T_{VOH1} = 0$, $T_{VIH0} = 0.20$, $T_{VIH1} = 0.05$, $T_{VIL0} = 0.15$, $T_{VIL1} = 0.05$, $T_{VOL0} = 0.02$, $T_{VOL1} = 0$.

Section II-C (and without loss of generality), we assume that $G^{(dr)}$ and $G^{(ld)}$ in a gate pair each represent a single logic stage. We also define the state of a logic stage input or output as its logic value (0 or 1). A logic stage input is sensitized if the logic stage output depends on the state of that input (given the logic values of all the other inputs).

For each logic stage input in our standard cell library, we model the VTC parameters for every case in which that input is sensitized (considering all possible combinations of the other inputs). The VTC parameters are functions of the CNT counts of the p- and n-type CNFETs (there is a CNT count variable for each CNFET in the circuit) which: 1) are gated by that input and 2) connect the logic stage output to either $V_{DD}$ or ground through a series of CNFETs in the "on" state (see [17, Sec. IX-B] for an example). We define $n_P$ as the sum of the CNT counts of all such p-type CNFETs. We similarly define $n_N$ for the n-type CNFETs. For example, inverter $U3$ in Fig. 8 consists of a p-type CNFET (labeled "$P3,1$") and an n-type CNFET ("$N3,1$"). The CNT counts of $P3,1$ and $N3,1$ are $n_P^{(P3,1)}$ and $n_N^{(N3,1)}$, respectively. Then $n_P = n_P^{(P3,1)}$ and $n_N = n_N^{(N3,1)}$.

For the NAND2 gate $U4$ in Fig. 8 (as an example of a logic stage with multiple inputs), we separately model the VTC parameters for each input: in1 and in2. Since there are two sets of VTC parameters for the NAND2 gate and only one output, the worst-case values for the output levels $V_{OH}$ and $V_{OL}$ (which are modeled as being independent of the CNT count: details at the end of this section) are selected from the two sets of VTC parameters (i.e., so that SNM is the lowest, see [17, Sec. IX-B] for an example).

Fig. 9 illustrates SPICE simulation data showing that the VTC of a logic stage is sensitive to CNT count variations. For example, the VTCs in Fig. 9 are representative of inverter $U3$ in Fig. 8, with CNFET CNT counts $n_P = n_P^{(P3,1)}$ and $n_N = n_N^{(N3,1)}$; variations in $n_P$ and $n_N$ cause variations in $V_{OH}$, $V_{IH}$, $V_{IL}$, and $V_{OL}$, resulting in SNM variations and larger PNMV.

We model the VTC parameters ($V_{OH}$, $V_{IH}$, $V_{IL}$, and $V_{OL}$) as affine functions of $\log_{10}(n_P/n_N)$ (this model is shown for an inverter in Fig. 9), which achieves a root-mean-square (RMS) modeling error $\leq 2.5$ mV in all cases (details in [17, Sec. IX-C]). For each case, this affine function is represented by a real-valued matrix $T \in \mathbb{R}^{4 \times 2}$

$$\begin{bmatrix} V_{OH} \\ V_{IH} \\ V_{IL} \\ V_{OL} \end{bmatrix} = T \begin{bmatrix} 1 \\ \log_{10}\left(\frac{n_P}{n_N}\right) \end{bmatrix}, T = \begin{bmatrix} T_{VOH0} & T_{VOH1} \\ T_{VIH0} & T_{VIH1} \\ T_{VIL0} & T_{VIL1} \\ T_{VOL0} & T_{VOL1} \end{bmatrix}. \quad (14)$$

To construct the full variation-aware SNM model (consisting of many instances of $T$ in our standard cell library: one for each combination of input states that sensitizes each logic stage input), we perform two steps for each instance of $T$.
1) Sample the CNT count for each CNFET in the logic stage 2000 times [53] (using the distribution of CNT count, given the CNFET widths and the experimentally demonstrated processing parameter values in Table I), and use SPICE simulations to obtain $V_{OUT}$ versus $V_{IN}$. For each sample, record $n_P$ and $n_N$ and extract $V_{OH}$, $V_{IH}$, $V_{IL}$, and $V_{OL}$ from the VTC in each simulation.
2) Find $T$ via linear regression, given the recorded $n_P$ and $n_N$ and the extracted $V_{OH}$, $V_{IH}$, $V_{IL}$, and $V_{OL}$.

We observed that in all cases, $T_{VOH1} \approx 0$ and $T_{VOL1} \approx 0$ (14), indicating that the CNT count ratio does not strongly affect the output levels of a logic stage.[2] Thus, to simplify our model, we set $T_{VOH1} = 0$ and $T_{VOL1} = 0$, and maintain RMS modeling error $\leq 2.5$ mV in all cases (details in [17, Sec. IX-C]). In Fig. 9, the VTC parameters are plotted versus $\log_{10}(n_P/n_N)$.

This variation-aware SNM model is critical for efficiently computing PNMV due to CNT count variations, as it relates the VTC parameters to the CNFET CNT counts for each logic stage (14). However, solving (5) for PNMV (in Section II-C) is not trivial due to CNT correlation (Section II-A), which causes correlated SNM among gate pairs. In Fig. 8, for example, gate pairs ($U1$, $U3$) and ($U3$, $U5$) have correlated SNM since the CNFETs in $U3$ and $U5$ have correlated CNT counts (they overlap the same sampling regions).

*2) Full-Circuit PNMV Model:* Here, we demonstrate how the variation-aware SNM model is used to efficiently calculate PNMV $\leq 0.001\%$ (which is desirable for VLSI-scale circuits: Section II-C) without using an MC-based technique (which would require many trials: e.g., $>10^5$ since $0.001\% = 1/10^5$). There are two key aspects in our framework for computing PNMV.
1) *PNMV Formulation:* We formulate PNMV as a function of the sampling region CNT count variables (which are independent) to account for the effects of CNT correlation (Section II-A) on SNM.
2) *Solving the PNMV Formulation Efficiently:* We provide a systematic technique to identify a small subset of all SNM constraints in the circuit [i.e., in (5) in Section II-C], referred to as the critical SNM constraints, which are the only SNM constraints that are required to compute PNMV. Due to CNT correlation, an SNM violation in a noncritical SNM constraint implies that there must also be an SNM violation in a critical SNM constraint; hence, the noncritical SNM constraint is not required to compute PNMV. For the OpenSPARC modules, <1% of all SNM constraints can be critical SNM constraints [17, Table VIII]. Hence, the time to compute PNMV (proportional to the number of SNM constraints) can improve by $>100\times$.

The first step to compute PNMV is to convert the SNM constraints in (5) into constraints on the CNT counts of each CNFET [using (2), (3), and (5) in Section II-C]. For each gate pair ($G^{(\text{dr})}$, $G^{(\text{ld})}$) [e.g., (*U3*, *U4*) in Fig. 8], each SNMH constraint has the form in (15) and each SNML constraint has the form in (16) (there can be multiple SNMH and SNML constraints for a single gate pair, details below)

$$\text{SNMH}\left(G^{(\text{dr})}, G^{(\text{ld})}\right) : V_{\text{OH}}^{(\text{dr})} - V_{\text{IH}}^{(\text{ld})} \geq SNM_R \quad (15)$$

$$\text{SNML}\left(G^{(\text{dr})}, G^{(\text{ld})}\right) : V_{\text{IL}}^{(\text{ld})} - V_{\text{OL}}^{(\text{dr})} \geq SNM_R. \quad (16)$$

---
[2]The effect of the CNT count ratio on the VTC is similar to that of the "beta ratio" $\beta_P/\beta_N$ (a measure of the relative strength of the pull-up and pull-down networks) in silicon-CMOS-based circuits, which does not have a strong effect on the output levels $V_{OH}$ and $V_{OL}$ [48].

We then substitute the variation-aware SNM model (14) into these constraints, using $T^{(\text{dr})}$ and $T^{(\text{ld})}$ to represent the SNM model for $G^{(\text{dr})}$ and $G^{(\text{ld})}$, respectively [e.g., $T^{(\text{dr})} = T^{(\text{INV\_X1})}$ and $T^{(\text{ld})} = T^{(\text{NAND2\_X1-in1})}$ for (*U3*, *U4*) in Fig. 8]

$$T_{\text{VOH0}}^{(\text{dr})} - T_{\text{VIH0}}^{(\text{ld})} - T_{\text{VIH1}}^{(\text{ld})} \log_{10}(n_P/n_N) \geq SNM_R \quad (17)$$

$$T_{\text{VIL0}}^{(\text{ld})} + T_{\text{VIL1}}^{(\text{ld})} \log_{10}(n_P/n_N) - T_{\text{VOL0}}^{(\text{dr})} \geq SNM_R. \quad (18)$$

These constraints are equivalently expressed in matrix form

$$\begin{bmatrix} 1 & \tilde{H}_{1,2} \\ \tilde{H}_{2,1} & 1 \end{bmatrix} \begin{bmatrix} n_P \\ n_N \end{bmatrix} \preceq \begin{bmatrix} 0 \\ 0 \end{bmatrix} \quad (19)$$

$$\tilde{H}_{1,2} = -10^{((T_{\text{VOH0}}^{(\text{dr})} - T_{\text{VIH0}}^{(\text{ld})} - SNM_R)/T_{\text{VIH1}}^{(\text{ld})})} \quad (20)$$

$$\tilde{H}_{2,1} = -10^{((T_{\text{VIL0}}^{(\text{ld})} - T_{\text{VOL0}}^{(\text{dr})} - SNM_R)/T_{\text{VIL1}}^{(\text{ld})})}. \quad (21)$$

Note that, the vector inequality in (19) is element-wise (as are all vector inequalities in this section). To account for all SNM constraints in the circuit, let $c$ be the total number of SNM constraints, and let $t$ be the total number of CNFETs (each with its own CNFET CNT count variable, e.g., $n_P^{(P3,1)}$ for CNFET *P3,1* in Fig. 8). For every gate pair ($G^{(\text{dr})}$, $G^{(\text{ld})}$), there is an SNMH constraint and an SNML constraint for each combination of input states that sensitizes the input to $G^{(\text{ld})}$ that is driven by $G^{(\text{dr})}$. For example, if $G^{(\text{dr})}$ drives an input of $G^{(\text{ld})}$ that can be sensitized by three combinations of input states [e.g., input "A" of an "and-or-invert" logic stage with Boolean function: out = (A + (B*C))$'$], then there are three SNMH and three SNML constraints for that gate pair (which may constrain different CNT count variables).

The total number of SNM constraints in the circuit is $c$, and each one imposes a constraint on the CNFET CNT count variables (e.g., $n_P^{(P3,1)}$, $n_N^{(N3,1)}$, $n_P^{(P4,1)}$, etc., in Fig. 8). We can represent these $c$ constraints with a single matrix inequality, by first defining a column vector $s \in \mathbb{R}^t$ that contains the CNT count variables for all the CNFETs in the circuit (e.g., if the entire circuit consisted of the ten CNFETs shown in Fig. 8, then $s = [n_P^{(P3,1)}; n_N^{(N3,1)}; n_P^{(P4,1)}; n_N^{(N4,1)}; n_P^{(P4,2)}; n_N^{(N4,2)}; n_P^{(P5,1)}; n_N^{(N5,1)}; n_P^{(P6,1)}; n_N^{(N6,1)}]$). Then by using all instances of $T$ in the SNM model (14), we formulate each SNM constraint as a constraint on the vector $s$, using the same procedure as above to convert (15)-(16) into constraints on the CNT counts in (19) (example in [17, Sec. IX-D]). We express these constraints using a matrix $H \in \mathbb{R}^{c \times t}$, such that satisfying (22) is equivalent to satisfying all SNM constraints in the circuit. [17, Table IX] summarizes all variables in this section.

$$Hs \preceq \mathbf{0}. \quad (22)$$

Note that, $\mathbf{0}$ is a column vector with element entry equal to 0. See [17, Sec. IX-D] for the formulation of (22) for the example circuit shown in Fig. 8. Since all SNM constraints in the circuit are represented in (22) (each of $c$ rows in $H$ represents a single SNM constraint), *PNMV* (5) is the probability that (22) is violated (i.e., $PNMV = 1 - P\{Hs \preceq \mathbf{0}\}$). However, solving for *PNMV* using (22) is not trivial since the CNT count variables (i.e., the elements of $s$) can be highly correlated due to CNT correlation (Section II-A). For example, in Fig. 8, the active regions of CNFETs *P3,1* and *P5,1* are aligned, so their CNT counts ($n_P^{(P3,1)}$ and $n_P^{(P5,1)}$) are correlated. Thus, the SNM constraints on $n_P^{(P3,1)}$ and the SNM constraints on $n_P^{(P5,1)}$ are dependent.

We can reformulate *PNMV* to efficiently account for CNT correlation by transforming the constraints in (22) (that constrain the CNFET CNT count variables, which are dependent) into constraints on the sampling region CNT count variables (which are independent). To do so, we first define a column vector $n \in \mathbb{R}^r$ that contains the CNT count variables for all the sampling regions (e.g., in Fig. 8, $n = [n_1; n_2; n_3; n_4; n_5; n_6; n_7; n_8; n_9; n_{10}; n_{11}; n_{12}; \ldots]$). To formulate (22) in terms of vector $n$ instead of vector $s$ ($s$: the CNFET CNT count variables), the relationship between $n$ and $s$ is required. We express this relationship as a linear transformation represented by a matrix $B \in \{0, 1\}^{t \times r}$ (details below) such that

$$s = Bn. \quad (23)$$

There is one row in $B$ for each CNFET in the circuit, and one column for each sampling region. To determine $B$: if CNFET $i$ (of $t$ total CNFETs) overlaps sampling region $j$ (of $r$ total sampling regions), then the value of $B$ in row $i$, column $j$ is 1 (i.e., $B_{i,j} = 1$); otherwise, $B_{i,j} = 0$ (as an example, $B$ for the circuit in Fig. 8 is shown in [17, Sec. IX-D]). Then by substituting (23) into (22), the SNM constraints can be expressed in terms of the region CNT count variables (instead of the CNFET CNT count variables), using a matrix $K \in \mathbb{R}^{c \times r}$

$$K = HB \quad (24)$$

$$Kn \preceq \mathbf{0}. \quad (25)$$

All SNM constraints in the circuit are represented in (25) [just as in (22)], so (25) can also be used to determine *PNMV* (i.e., $PNMV = 1 - P\{Kn \preceq \mathbf{0}\}$). The advantage of using (25) instead of (22) is that all the variables in $n$ (the vector of sampling region CNT counts) are independent (unlike the correlated variables in $s$: the vector of CNFET CNT counts). For example, (25) can be used to estimate *PNMV* via an MC-based approach: for each trial, sample all elements of $n$ (from the distribution of sampling region CNT count) and evaluate $Kn$. Then estimate *PNMV* as the fraction of trials that violate (25).

However, evaluating every SNM constraint in (25) is unnecessary since many of them are noncritical SNM constraints (as described above, any SNM constraint that cannot be uniquely violated without simultaneously violating another SNM constraint is not required to compute *PNMV*). See [17, Sec. IX-E] for a detailed description, including examples, of how to systematically identify and eliminate all noncritical SNM constraints.

Eliminating these noncritical SNM constraints is crucial to improve computational efficiency, as they can account for $\geq 99\%$ of all SNM constraints in the circuit (e.g., for the OpenSPARC modules [17, Table VIII]). Since each row of $K$ (25) represents an SNM constraint, we can remove the rows in $K$ that correspond to noncritical SNM constraints to form $\tilde{K} \in \mathbb{R}^{p \times r}$, where $p$ is the number of critical SNM constraints

$$\tilde{K}n \preceq \mathbf{0}. \quad (26)$$

To further improve computational efficiency, we then factor out $\mu_R$ and $\sigma_R$ from the sampling region CNT count variables $n$ in (26) (just as we did for the full-circuit delay model in Section III-A), allowing us to quickly recompute *PNMV* after updating the processing parameter values (details in Section III-D). We rewrite $n = \mu_R \mathbf{1} + \sigma_R x$, where each entry of $x \in \mathbb{R}^r$ is distributed according to a unit Gaussian distribution; then (26) becomes

$$\tilde{K}x \preceq (\mu_R/\sigma_R)b \quad (27)$$
$$b = -\tilde{K}\mathbf{1}. \quad (28)$$

In (27), $b$ is a $p$-dimensional vector of constants and the matrix-vector product $\tilde{K}x$ is a $p$-dimensional vector of Gaussian random variables with covariance matrix $C \in \mathbb{R}^{p \times p}$

$$C = \tilde{K}\tilde{K}^T. \quad (29)$$

That is, $\tilde{K}x$ is distributed according to a multivariate normal (MVN) distribution with covariance matrix $C$; thus, $PNMV = 1 - P\{\tilde{K}x \preceq (\mu_R/\sigma_R)b\}$ can be solved numerically using existing software packages for computing MVN probabilities (see [14]). In particular, consider the cumulative distribution function (CDF) of the MVN-distributed matrix-vector product $\tilde{K}x$ (i.e., the MVNCDF); the probability that all SNM constraints are satisfied (i.e., $1 - PNMV$) is equal to the value of the MVNCDF at the $p$-dimensional point $(\mu_R/\sigma_R)b$

$$PNMV = 1 - P\{\tilde{K}x \preceq (\mu_R/\sigma_R)b\} \quad (30)$$
$$PNMV = 1 - \text{MVNCDF}(C, (\mu_R/\sigma_R)b). \quad (31)$$

In [17, Sec. IX-F], we describe how to efficiently solve (31) by leveraging the property that many terms in the covariance matrix, $C$, are 0; e.g., for the OpenSPARC modules, *PNMV* can be computed in less than 10 seconds using a single 2.93 GHz processor core. In [17, Fig. 21], we validate the accuracy of (31) against MC simulations. For the MC approach, we first sample the vector of sampling region CNT counts [$n$ in (25)] for each trial. Then we estimate *PNMV* as the fraction of samples that violate (25).

### C. Circuit Performance Sensitivity to Processing Parameters

Our goal is to achieve small delay penalties and *PNMV* with small $\Delta E$. We quantify the tradeoff between total circuit energy [$E_{\text{Tot}}$ in (13)] and delay penalty using $EDP_{95}$: defined in (32). We also define the energy-*PNMV*-product (*ENP*) metric to quantify the tradeoff between $E_{\text{Tot}}$ and *PNMV*

$$EDP_{95} = E_{\text{Tot}}T_{95} \quad (32)$$
$$ENP = E_{\text{Tot}}PNMV. \quad (33)$$

While rapid computation of circuit delay penalty and *PNMV* overcomes the computation time bottleneck of analyzing a single design point, we still require a method for intelligently exploring the large space of CNT processing options. In general, a common measure of the sensitivity of an objective function (e.g., $EDP_{95}$ or *ENP*) with respect to each of its input variables (e.g., the processing parameters) is its gradient. The $EDP_{95}$ and *ENP* gradients are defined in (34)–(36) and are used to guide the exploration of processing options to improve delay penalties and *PNMV* (Section III-D). Fig. 10 illustrates a flowchart of the steps used to compute (32)–(36). The gradients (34)–(36) are computed as described [17, Sec. X-A].

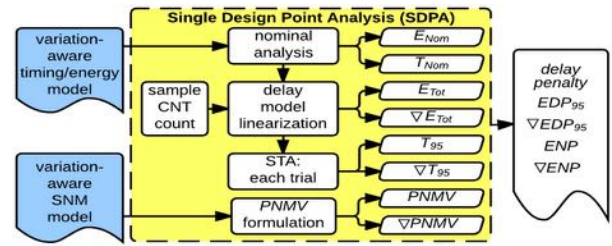

Fig. 10. SDPA to calculate the delay penalty, $EDP_{95}$, $\nabla EDP_{95}$, *ENP*, and $\nabla ENP$ of a single design point.

$$\nabla E_{\text{Tot}} = \left[\frac{\partial E_{\text{Tot}}}{\partial IDC}; \frac{\partial E_{\text{Tot}}}{\partial p_m}; \frac{\partial E_{\text{Tot}}}{\partial p_{\text{Rs}}}\right]$$
$$\nabla T_{95} = \left[\frac{\partial T_{95}}{\partial IDC}; \frac{\partial T_{95}}{\partial p_m}; \frac{\partial T_{95}}{\partial p_{\text{Rs}}}\right] \quad (34)$$
$$\nabla PNMV = \left[\frac{\partial PNMV}{\partial IDC}; \frac{\partial PNMV}{\partial p_m}; \frac{\partial PNMV}{\partial p_{\text{Rs}}}\right]$$
$$\nabla EDP_{95} = \nabla E_{\text{Tot}}T_{95} + E_{\text{Tot}}\nabla T_{95} \quad (35)$$
$$\nabla ENP = \nabla E_{\text{Tot}}PNMV + E_{\text{Tot}}\nabla PNMV. \quad (36)$$

### D. Guided Exploration to Overcome CNT Variations

To overcome the bottleneck of trial-and-error-based search (i.e., iterating over many combinations of values for the processing parameters and design parameters defined in Section II: *IDC*, $p_m$, $p_{\text{Rs}}$, $p_{\text{Rm}}$, $k_{\text{SelUpsize}}$), we use a gradient descent-based strategy to systematically guide the improvement of $EDP_{95}$ and *ENP* in the presence of CNT variations (while gradient descent strategies can converge to local rather than global optima, in [17, Sec. X-C] we discuss techniques to reduce the impact of local optima during gradient descent in our methodology).

For any design point, we can use single design point analysis (SDPA: Fig. 10) to determine the sensitivity of each circuit performance metric (e.g., $EDP_{95}$ or *ENP*) to each processing parameter by computing its gradient (e.g., $\nabla EDP_{95}$ or $\nabla ENP$). These gradients can then be used to identify which processing parameters should be improved (and by how much) to efficiently improve the circuit performance metrics. For example, consider $EDP_{95}$: $\partial EDP_{95}/\partial IDC$, $\partial EDP_{95}/\partial p_m$, and $\partial EDP_{95}/\partial p_{\text{Rs}}$ indicate how sensitive $EDP_{95}$ is to improvements in *IDC*, $p_m$, and $p_{\text{Rs}}$. Thus, to effectively improve $EDP_{95}$, we can update each processing parameter by an amount proportional to its corresponding value in $\nabla EDP_{95}$ (e.g., *IDC* corresponds to $\partial EDP_{95}/\partial IDC$). We refer to each such update as a gradient descent step (details in [17, Sec. X-B]).

Before describing the full gradient descent methodology, we define the initial design point as the design point after EDP optimization in the nominal case (i.e., after steps 1–3 in Fig. 3). Also, we define the initial processing parameter values as the processing parameter values of the initial design point (e.g., $IDC = 0.50$, $p_m = 10\%$, $p_{\text{Rs}} = 4\%$, $p_{\text{Rm}} = 99.99\%$: Table I). Starting from the initial design point, we first perform selective upsizing (as described in Section II-A), incrementally increasing $k_{\text{SelUpsize}}$ (the number of standard cells to upsize) to generate a set of design points referred to as the initial energy-delay tradeoff curve (the values of $k_{\text{SelUpsize}}$ are chosen so that each increase in $k_{\text{SelUpsize}}$ increases $\Delta E$ by ~1%–2% to identify multiple design points with $\Delta E \leq 5\%$).

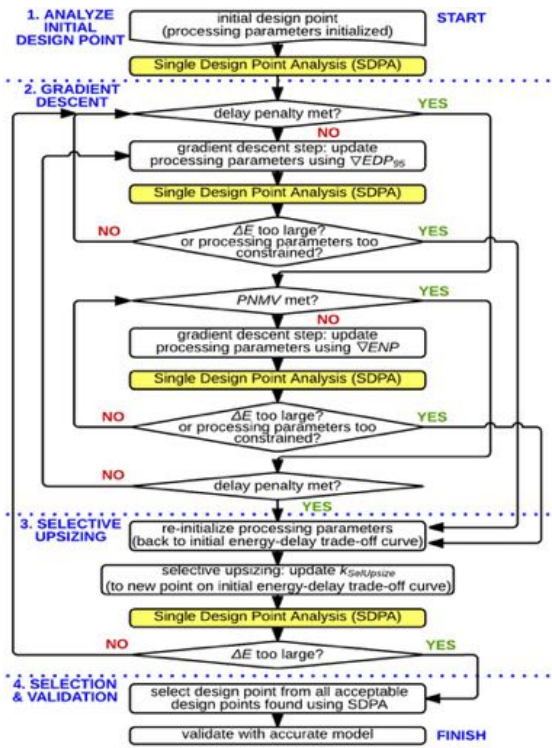

Fig. 11. Gradient descent-based methodology to meet delay penalty, *PNMV*, and $\Delta E$ requirements. SDPA details in Fig. 10.

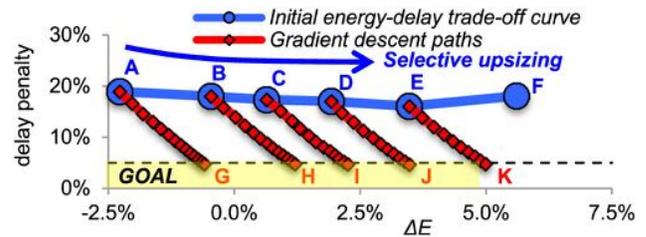

Fig. 12. Gradient descent methodology (Fig. 11) to achieve delay penalty $\leq 5\%$, $PNMV \leq 0.001\%$ (for $SNM_R = V_{DD}/6$), $\Delta E \leq 5\%$ (5 nm "pku" OpenSPARC module). Gradient descent paths descend from the initial energy-delay tradeoff curve ($IDC = 0.50$, $p_m = 10\%$, $p_{Rs} = 4\%$, $p_{Rm} = 99.99\%$). The point (delay penalty, $\Delta E$) = (0%, 0%) represents the EDP-optimized nominal design point (Section II-A, Fig. 3). $\Delta E < 0$ for point A since $E_{Tot}$ depends on the number of s-CNTs after m-CNT removal (i.e., the CNT count variables), as shown in (13) in Section III-A; due to CNT count variations (e.g., resulting from $p_m > 0\%$, $p_{Rs} > 0\%$), the total number of s-CNTs in all CNFETs can be reduced versus the nominal case (no variations).

The full methodology, illustrated in Fig. 11, combines selective upsizing and gradient descent to overcome the impact of CNT count variations on delay penalty and *PNMV*. After generating the initial energy-delay tradeoff curve via selective upsizing, our goal is to identify *multiple* design points that meet both a delay penalty constraint (e.g., delay penalty $\leq 5\%$) and a *PNMV* constraint (e.g., $PNMV \leq 0.001\%$) with minimal energy cost (e.g., $\Delta E \leq 5\%$). Such design points that simultaneously satisfy all these design goals are referred to as acceptable design points. Consequently, this is a feasibility problem in which we search for design points that meet two constraints, and we solve it using a variation of an alternating projections (AP) algorithm [3]. A typical AP algorithm iteratively projects a point onto multiple constraints until all are satisfied. In our methodology, we use gradient descent instead of projection; the full methodology (Fig. 11) is described below (example in Fig. 12).

1) *Analyze the Initial Design Point:* Perform SDPA (Fig. 10) on the initial design point [with initial processing parameter values and $k_{SelUpsize}$ set to minimize $EDP_{NomOpt}$ (1): i.e., after steps 1–3 in Fig. 3].
2) *Gradient Descent:* Alternate between: 1) performing gradient descent steps using $\nabla EDP_{95}$ until the delay penalty constraint is satisfied and 2) performing gradient descent steps using $\nabla ENP$ until the *PNMV* constraint is satisfied. This procedure continues until either: a) both constraints are satisfied simultaneously (i.e., an acceptable design point is found) or b) $\Delta E$ is too large or the processing parameters are too constrained (e.g., a design point with $\Delta E > 5\%$ is reached, or the required processing parameter values may be difficult to achieve experimentally: both are design choices).
3) *Selective Upsizing:* Reinitialize the processing parameters to their initial values (thus returning to the initial energy-delay tradeoff curve) and then perform selective upsizing (by increasing $k_{SelUpsize}$) to move to the next point on the initial energy-delay tradeoff curve. If $\Delta E$ from selective upsizing is too large (e.g., $\Delta E > 5\%$), then proceed to step 4 (below). Otherwise, loop back to step 2 (gradient descent) to search for an additional acceptable design point.
4) *Design Point Selection and Validation:* Select a single design point from all acceptable design points identified using gradient descent. For example, the designer can select the acceptable design point with the minimum $EDP_{95}$ or with the most relaxed processing requirements (a design choice). Finally, highly accurate models (e.g., the nonlinear timing model) can be used to validate the selected design point (if all constraints are not satisfied during validation, then perform additional gradient descent steps until they are satisfied).

Fig. 12 illustrates an example of the gradient descent-based methodology (Fig. 11) to meet delay penalty $\leq 5\%$ and *PNMV* $\leq 0.001\%$ with $\Delta E \leq 5\%$. Starting from point A (the initial design point), we perform selective upsizing to generate the initial energy-delay tradeoff curve (as described earlier in this section) represented by points A–F. Then, using the methodology in Fig. 11, we perform gradient descent (starting from the initial design point: point A) until delay penalty $\leq 5\%$ and $PNMV \leq 0.001\%$ (at point G: an acceptable design point). Next, the processing parameters are reinitialized and then selective upsizing brings us to point B on the initial energy-delay tradeoff curve. Again, gradient descent is performed to identify another acceptable design point (point H). This process repeats until we reach point F on the initial energy-delay tradeoff curve, which has $\Delta E > 5\%$, concluding the search for acceptable design points.

In Fig. 12, gradient descent has identified *multiple* acceptable design points with varying $\Delta E$ and processing requirements. Furthermore, alternative sets of acceptable design points can be identified by adjusting the gradient descent step procedure: e.g., if *IDC* is difficult to improve (i.e., it is difficult to control CNT density variations experimentally), then the gradient descent step can be weighted toward larger updates in $p_m$ or $p_{Rs}$, or can be forced never to update *IDC* past a predetermined hard-limit. These constraints can be provided as inputs, and are features of this flexible framework.

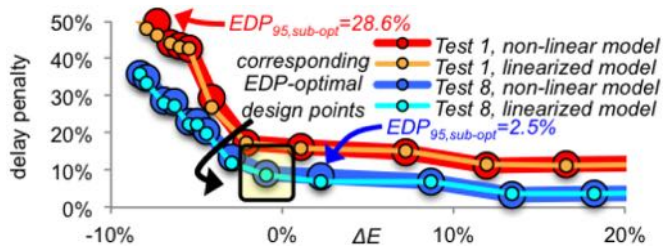

Fig. 13. Delay penalty vs. $\Delta E$ (5 nm "gkt" OpenSPARC module). Large markers: values computed using the nonlinear model; small markers: linearized model. Processing parameter values are in [17, Table VII] (tests 1 & 8). Both models identify that the same design point has the minimum $EDP_{95}$, thus $EDP_{95,\text{sub-opt}} = 0\%$. If the models had selected different design points [e.g., if the linearized model had selected the labeled point at (2.3%, 8.2%) with $EDP_{95,\text{sub-opt}} = 2.5\%$] then $EDP_{95,\text{sub-opt}} \geq 0\%$. The point (delay penalty, $\Delta E$) = (0%, 0%) represents the EDP-optimized nominal design point (Section II-A, Fig. 3). Design points with $\Delta E < 0$ are for designs with smaller CNFET widths.

## IV. RESULTS

We present two sets of results to demonstrate that we have overcome the bottlenecks of brute-force trial-and-error-based approaches. The first set of results (Section IV-A) demonstrates that we can analyze a set of design points >100× faster than before, while maintaining sufficient accuracy to make correct design decisions. The second set of results (Section IV-B) demonstrates the ability of this gradient descent algorithm to identify multiple processing options to meet design goals (e.g., delay penalty ≤5% and $PNMV \leq 0.001\%$ with $\Delta E \leq 5\%$) without exhaustive search. Using the results from gradient descent, we provide practical processing guidelines for each node and for multiple values of $V_{DD}$ such that: even in the presence of CNT variations, CNFET circuits can maintain ≥90% of the projected EDP benefits of nominal CNFET circuits.

### A. Linearized Timing Model Validation

Here, we validate the speed and accuracy of the linearized timing model to analyze circuit delay variations. We first choose a set of design points that would typically be chosen by a designer seeking to optimize $EDP_{95}$ using a brute-force-search-based approach: we use the design points in [52] as a reference. We analyze 112 design points including all combinations of: eight unique sets of processing parameter values (the same as in [52]: see [17, Table VII]), and 14 different $k_{\text{SelUpsize}}$ values (each increase in $k_{\text{SelUpsize}}$ increases $\Delta E$ by ~1%–5%, e.g., in Fig. 13; higher resolution requires more computation time).

After choosing the set of design points, we use the nonlinear timing model to compute $EDP_{95}$ of every design point and then select the design point with the best (minimum) $EDP_{95}$. We also record the total required computation time. We then perform the same procedure using the linearized timing model. We evaluate: 1) the total computation time for each model and 2) the degradation (increase) in $EDP_{95}$ due to using the linearized timing model. To quantify this degradation, we use the $EDP_{95}$ sub-optimality metric defined in (37); it is computed using only the nonlinear timing model to compare: $EDP_{95}$ of the design point selected by each model

$$EDP_{95,\text{sub-opt}} = \frac{EDP_{95,\text{Measured By NonLinear Model}}^{(\text{Selected By Linearized Model})}}{EDP_{95,\text{Measured By NonLinear Model}}^{(\text{Selected By NonLinear Model})}} - 1. \quad (37)$$

TABLE III
$EDP_{95}$ SUB-OPTIMALITY AND COMPUTATION TIME (MEASURED ON A SINGLE 2.93 GHZ PROCESSOR WITH NO PARALLELIZATION). LOGIC GATE COUNT IS TAKEN FROM THE SYNTHESIZED NETLIST AT THE 5 NM NODE (WITH $V_{DD} = 0.50$ V)

| Open-SPARC module | Logic gate count | Time, non-linear timing model | Time, linearized timing model | Speed -up | $EDP_{95}$ sub-optimality: $EDP_{95,\text{sub-opt}}$ |
|---|---|---|---|---|---|
| dec | 4.0K | 1.6 days | 20 minutes | 112X | 0% |
| pmu | 9.9K | 3.4 days | 40 minutes | 121X | 0% |
| pku | 10.7K | 4.1 days | 50 minutes | 119X | 1.5% |
| gkt | 10.6K | 3.6 days | 41 minutes | 126X | 0% |
| exu | 19.5K | 1 week | 1.3 hours | 124X | 0% |
| lsu | 46.5K | 3 weeks | 3.2 hours | 160X | 0% |
| tlu | 69.5K | 1 month | 4.7 hours | 162X | 0.9% |
| fgu | 104.1K | 1.4 months | 6.0 hours | 172X | 0.3% |

TABLE IV
PROCESSING ROUTES TO MEET DELAY PENALTY CONSTRAINTS AND $PNMV \leq 0.001\%$ WITH $\Delta E \leq 5\%$ FOR ALL OPENSPARC MODULES SIMULTANEOUSLY. IN ALL CASES: $V_{DD} = 0.50$ V, $SNM_R = V_{DD}/6$, $p_{Rm} = 99.99\%$, COUNT-LIMITED YIELD ≥99.999%

| Delay penalty | 5 nm node: $IDC, p_m, p_{Rs}$ | 7 nm node: $IDC, p_m, p_{Rs}$ | 10 nm node: $IDC, p_m, p_{Rs}$ | 14 nm node: $IDC, p_m, p_{Rs}$ |
|---|---|---|---|---|
| ≤ 5% | 0.19, 0.8%, 2.2% | 0.20, 0.9%, 2.3% | 0.19, 0.9%, 2.6% | 0.23, 0.9%, 2.6% |
| ≤ 6% | 0.25, 0.9%, 2.6% | 0.26, 0.9%, 2.6% | 0.26, 0.9%, 2.8% | 0.31, 0.9%, 2.9% |
| ≤ 7% | 0.30, 0.9%, 2.7% | 0.31, 0.9%, 2.8% | 0.32, 0.9%, 2.9% | *0.35, 0.9%, 3.0%* |
| ≤ 8% | *0.31, 0.9%, 2.7%* | *0.32, 0.9%, 2.8%* | *0.35, 0.9%, 2.9%* | *0.35, 0.9%, 3.0%* |

Ideally, the same design point is selected using each of the two models (resulting in $EDP_{95,\text{sub-opt}} = 0\%$, example in Fig. 13). This is the case for five of the eight OpenSPARC modules (5 nm node), and the other three have $EDP_{95,\text{sub-opt}} \leq 2\%$ (Table III).[3] The linearized model achieves >100× speed-up in all cases.

### B. CNT Processing and CNFET Circuit Design Guidelines

We now demonstrate the effectiveness of the gradient descent methodology to identify multiple sets of guidelines for processing parameters (i.e., processing routes) that meet design goals for all OpenSPARC modules simultaneously. For each OpenSPARC module, we first perform gradient descent (Fig. 11, with initial processing parameter values: $IDC = 0.50$, $p_m = 1\%$, $p_{Rs} = 4\%$, $p_{Rm} = 99.99\%$) to identify multiple acceptable design points (with delay penalty ≤5%, $PNMV \leq 0.001\%$, $\Delta E \leq 5\%$), and then we select the design point with the most relaxed processing requirements (though other selection criteria can be used, e.g., lowest $EDP_{95}$). Then, for each processing parameter, we select its most constrained value (i.e., the value closest to its ideal value: Table I) over all the selected design points (one for each OpenSPARC module). These values form a processing route, and we then validate that design goals are met for all modules for this processing route (e.g., using the nonlinear model to compute delay penalty). Table IV provides processing routes for the OpenSPARC modules at the 14, 10, 7, and 5 nm nodes (highlighted entries in Table IV are limited by the $PNMV$ constraint; other entries are limited by the delay penalty constraint). For each node, processing routes are shown for multiple delay penalty constraints to illustrate the tradeoff between delay penalty, $PNMV$, and processing requirements. All processing routes in Table IV meet count-limited yield ≥99.999%, resulting from minimum-width upsizing (step 2 in Fig. 3: to reach count-limited yield ≥99.9%) and CNT process improvements; if count-limited yield <99.999%, then we can return to step 2 in Fig. 3 to increase $W_{MIN}$, then repeat gradient descent (Fig. 11) to find processing routes.

---
[3]In general, $EDP_{95,\text{sub-opt}}$ depends on the chosen set of design points since there is a finite number of possible values for $EDP_{95,\text{sub-opt}}$; results in Table III reflect a typical brute-force-based EDP optimization [52].

TABLE V
PROCESSING ROUTES TO MAINTAIN EDP BENEFIT ≥90% (VERSUS THE NOMINAL CASE) AND $PNMV \leq 0.001\%$ ($SNM_R = V_{DD}/6$) WITH $\Delta E \leq 5\%$ (ALL OPENSPARC MODULES). COUNT-LIMITED YIELD ≥99.999% IN ALL CASES

| Node | $V_{DD}$ | $IDC$ | $p_m$ | $p_{Rs}$ | $p_{Rm}$ |
|---|---|---|---|---|---|
| 14 nm | 0.50 V | 0.35 | 0.9% | 3.0% | 99.99% |
| 10 nm | 0.50 V | 0.33 | 0.9% | 2.9% | 99.99% |
| 7 nm | 0.50 V | 0.31 | 0.9% | 2.8% | 99.99% |
| 5 nm | 0.50 V | 0.30 | 0.9% | 2.7% | 99.99% |
| 5 nm | 0.45 V | 0.29 | 0.9% | 2.6% | 99.99% |
| 5 nm | 0.40 V | 0.27 | 0.9% | 2.6% | 99.99% |
| 5 nm | 0.35 V | 0.25 | 0.9% | 2.5% | 99.99% |

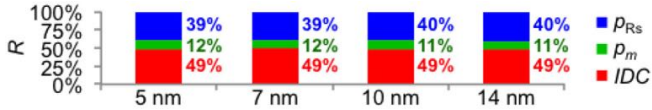

Fig. 14. Relative improvement of each processing parameter (versus $IDC = 0.50$, $p_m = 1\%$, $p_{Rs} = 4\%$) for processing routes in Table V ($V_{DD} = 0.50$ V). $p_{Rm}$ is not shown since $p_{Rm} = 99.99\%$ in all cases in Table V.

We have so far targeted delay penalty ≤5%, $PNMV \leq 0.001\%$, and $\Delta E \leq 5\%$, which maintains ≥90% of the projected EDP benefits of nominal CNFET circuits despite CNT variations. However, achieving these design goals can impose processing requirements that may be difficult to achieve experimentally (e.g., $IDC = 0.19$ for delay penalty ≤5% at the 5 nm node: Table IV). In Table V, we provide alternative processing routes that maintain ≥90% of the projected EDP benefits of nominal CNFET circuits; we target design points with EDP benefit ≥90% (versus nominal) with a relaxed delay penalty constraint (≤10%, resulting in lower $\Delta E$ to meet the EDP benefit goal).

The amount by which each processing parameter is improved is a measure of its effectiveness to improve delay penalty and *PNMV* (gradient descent incurs larger updates for processing parameters that more significantly impact these performance metrics, details in [17, Sec. X-B]). Fig. 14 shows the relative improvement [$R$ in (40)] of $IDC$, $p_m$, and $p_{Rs}$ from their initial values to their final values (in Table V). $R$ is calculated using the percentage improvement ($I$) and the total improvement ($I_{Tot}$) of the processing parameters

$$(I_{IDC}, I_{pm}, I_{pRs}) = \left(1 - \frac{IDC^{final}}{IDC^{init}}, 1 - \frac{p_m^{final}}{p_m^{init}}, 1 - \frac{p_{Rs}^{final}}{p_{Rs}^{init}}\right) \quad (38)$$

$$I_{Tot} = I_{IDC} + I_{pm} + I_{pRs} \quad (39)$$

$$(R_{IDC}, R_{pm}, R_{pRs}) = \left(\frac{I_{IDC}}{I_{Tot}}, \frac{I_{pm}}{I_{Tot}}, \frac{I_{pRs}}{I_{Tot}}\right). \quad (40)$$

The relative improvement is highest for $IDC$ for all nodes, showing that $IDC$ is a highly effective parameter to improve for reducing delay penalties and *PNMV* in an energy-efficient manner. From our results, we make the following conclusions.
1) The computationally efficient linearized timing model runs over 100× faster than the nonlinear timing model, and maintains sufficient accuracy to identify design points with $EDP_{95,sub-opt} \leq 2\%$ for all test cases.
2) $PNMV \leq 0.001\%$ can be efficiently computed.
3) Gradient descent is a systematic and scalable method to meet both delay penalty and *PNMV* constraints.
4) Gradient descent can efficiently identify multiple processing routes to meet design goals.
5) In contrast to traditional thinking (which focuses on reducing $p_m$ to ultralow values), gradient descent identifies that reducing $IDC$ is a highly effective means of meeting delay penalty and *PNMV* constraints, and that reducing $p_m$ past 1% suffers from diminishing returns. Unlike trial-and-error approaches [52], gradient descent establishes these facts in a highly rigorous manner.

## V. CONCLUSION

We have demonstrated a systematic methodology for joint exploration and optimization of CNT processing and CNFET circuit design to overcome the significant challenge of CNT variations. Our approach enables quick evaluation of delay variations and *PNMV* of CNFET VLSI circuits with >100× speed-up versus existing approaches. Our gradient descent-based framework accurately identifies the most important processing parameters, in conjunction with CNFET circuit sizing, to achieve high energy efficiency while satisfying circuit-level noise margin and yield constraints. Using this framework, an important question regarding CNT variations can be answered.

*Question:* What values of $IDC$, $p_m$, $p_{Rs}$, and $p_{Rm}$ should be targeted for highly scaled VLSI CNFET circuits to maintain a significant portion of their projected speed and energy efficiency benefits despite CNT variations, while also meeting circuit-level noise margin and yield constraints?

*Answer:* At the 5 nm node, we recommend $IDC = 0.25$, $p_m = 0.9\%$, $p_{Rs} = 2.5\%$, and $p_{Rm} = 99.99\%$ to maintain ≥90% of the projected EDP benefits versus nominal CNFET circuits, with $PNMV \leq 0.001\%$, functional yield ≥99.999%, and $\Delta E \leq 5\%$. These processing guidelines are attractive since $p_m = 1\%$ and $p_{Rm} = 99.99\%$ have been experimentally demonstrated, $p_{Rs} = 4\%$ has been achieved, and promising work for continued improvement of $p_{Rs}$ has been shown [19]. This leaves $IDC$ to be improved by 2× (versus $IDC = 0.50$: shown experimentally), thus identifying CNT density variations as an important topic of research. Additionally, processing requirements may be further relaxed by combining various CNT processing techniques (e.g., CNT sorting [1] followed by VMR [32]). Processing routes for other nodes are provided in Table V.

Unlike existing trial-and-error techniques, our framework can systematically explore the large space of CNT processing options, and generate a variety of processing routes depending on CNT processing technology constraints. Such systematic exploration is essential for a successful CNFET technology to avoid potential obstacles. Future research directions include the following.
1) Incorporation of CNT-metal contact resistance variations and threshold voltage variations into our framework, as well as other CNT processing techniques (e.g., [19]).
2) Experimental validation of model parameters for high-density CNT growth techniques and for channel lengths closer to the ballistic regime [41], [43].
3) Examination of the applicability of our framework for other emerging nanotechnologies, as many emerging nanotechnologies are expected to exhibit substantial variations. Our methodology can be adapted to overcome challenges in those technologies as well.

## ACKNOWLEDGMENT

The authors would like to thank STARnet SONIC, NSF, the Stanford Graduate Fellowship, and the Hertz Foundation

Fellowship (for M. M. Shulaker). They would also like to thank S. Keller, California Institute of Technology, S. Sinha, ARM, Inc., and Prof. D. Antoniadis, Massachusetts Institute of Technology, for valuable discussions.

## REFERENCES

[1] M. S. Arnold, A. A. Green, J. F. Hulvat, S. I. Stupp, and M. C. Hersam, "Sorting carbon nanotubes by electronic structure using density differentiation," *Nat. Nanotech.*, vol. 1, no. 1, pp. 60–65, 2006.

[2] S. Bangsaruntin, G. M. Cohen, A. Majumdar, and J. W. Sleight, "Universality of short-channel effects in undoped-body silicon nanowire MOSFETs," *IEEE Electron Device Lett.*, vol. 31, no. 9, pp. 903–905, Sep. 2010.

[3] H. H. Bauschke and J. M. Borwein, "Dykstra's alternating projection algorithm for two sets," *J. Approx. Theory*, vol. 79, no. 3, pp. 418–443, 1994.

[4] G. Budiman, Y. Gao, X. Wang, S. Koswatta, and M. Lundstrom. (2014). *Cylindrical CNTMOSFET Simulator*. [Online]. Available: https://nanohub.org/resources/moscntr

[5] Q. Cao et al., "Medium-scale carbon nanotube thin-film integrated circuits on flexible plastic substrates," *Nature*, vol. 454, pp. 495–500, Jul. 2008.

[6] L. Chang et al., "IEDM short course," in *Proc. Int. Electron Device Meeting (IEDM)*, San Francisco, CA, USA, 2012.

[7] Z. Chen et al., "An integrated logic circuit assembled on a single carbon nanotube," *Science*, vol. 311, no. 5768, p. 1735, 2006.

[8] C. Chu, "FLUTE: Fast lookup table based wirelength estimation technique," in *Proc. Int. Conf. Comput.-Aided Design (ICCAD)*, San Jose, CA, USA, 2004, pp. 696–701.

[9] J. Deng et al., "Carbon nanotube transistor circuits: Circuit-level performance benchmarking and design options for living with imperfections," in *Proc. Int. Solid-State Circuits Conf. (ISSCC)*, San Francisco, CA, USA, 2007, pp. 70–588.

[10] R. Dennard et al., "Design of ion-implanted MOSFET's with very small physical dimensions," *IEEE J. Solid-State Circuits*, vol. SC-9, no. 5, pp. 256–268, Oct. 1974.

[11] L. Ding et al., "CMOS-based carbon nanotube pass-transistor logic integrated circuits," *Nat. Commun.*, vol. 3, Feb. 2012, Art. ID 677.

[12] (Jul. 2012). *Eigen C++ Template Library for Linear Algebra*. [Online]. Available: http://eigen.tuxfamily.org

[13] A. D. Franklin et al., "Sub-10 nm carbon nanotube transistor," *Nano Lett.*, vol. 12, no. 2, pp. 758–762, 2012.

[14] A. Genz, "Numerical computation of multivariate normal probabilities," *J. Comput. Graph. Stat.*, vol. 1, no. 2, pp. 141–149, 1992.

[15] M. C. Hansen et al., "Unveiling the ISCAS-85 benchmarks: A case study in reverse engineering," *IEEE Des. Test*, vol. 16, no. 3, pp. 72–80, Jul. 1999.

[16] G. Hills et al., "Rapid exploration of processing and design guidelines to overcome carbon nanotube variations," in *Proc. Design Autom. Conf.*, Austin, TX, USA, 2013, pp. 1–10.

[17] G. Hills et al. *Rapid Co-Optimization of Processing and Circuit Design to Overcome Carbon Nanotube Variations*. [Online]. Available: http://www.arxiv.org

[18] (2013). *ITRS*. [Online]. Available: http://www.itrs.net/Links/2013ITRS/Home2013.htm

[19] S. H. Jin et al., "Using nanoscale thermocapillary flows to create arrays of purely semiconducting single-walled carbon nanotubes," *Nat. Nanotechnol.*, vol. 8, no. 5, pp. 347–355, 2013.

[20] S. J. Kang et al., "High-performance electronics using dense, perfectly aligned arrays of single-walled carbon nanotubes," *Nat. Nanotechnol.*, vol. 2, no. 4, pp. 230–236, 2007.

[21] A. Khakifirooz et al., "A simple semiempirical short-channel MOSFET current–voltage model continuous across all regions of operation and employing only physical parameters," *IEEE Trans. Electron Devices*, vol. 56, no. 8, pp. 1674–1680, Aug. 2009.

[22] J. Lohstroh et al., "Worst-case static noise margin criteria for logic circuits and their mathematical equivalence," *IEEE J. Solid-State Circuits*, vol. 18, no. 6, pp. 803–807, Dec. 1983.

[23] J. Luo et al., "A compact model for carbon nanotube field-effect transistors including non-idealities and calibrated with experimental data down to 9 nm gate length," *IEEE Trans. Electron Devices*, vol. 60, no. 6, pp. 1834–1843, Jun. 2013.

[24] D. Markovic, V. Stojanovic, B. Nikolic, M. A. Horowitz, and R. W. Brodersen, "Methods for true energy-performance optimization," *IEEE J. Solid-State Circuits*, vol. 39, no. 8, pp. 1282–1293, Aug. 2004.

[25] (Aug. 2014). *Nangate Open Cell Libraries*. [Online]. Available: http://www.nangate.com

[26] S. Nassif et al., "High performance CMOS variability in the 65 nm regime and beyond," in *Proc. IEEE Int. Electron Devices Meeting (IEDM)*, Washington, DC, USA, 2007, pp. 569–571.

[27] (Dec. 2011). *OpenSPARC*. [Online]. Available: http://www.opensparc.net/opensparc-t2

[28] H. Park et al., "High-density integration of carbon nanotubes via chemical self-assembly," *Nat. Nanotechnol.*, vol. 7, no. 12, pp. 787–791, 2012.

[29] J. Parker, C. Beasley, A. Lin, H.-Y. Chen, and H.-S. P. Wong, "Increasing the semiconducting fraction in ensembles of single-walled carbon nanotubes," *Carbon*, vol. 50, no. 14, pp. 5093–5098, 2012.

[30] N. Patil, J. Deng, A. Lin, H.-S. P. Wong, and S. Mitra, "Design methods for misaligned and mispositioned carbon-nanotube immune circuits," *IEEE Trans. Comput.-Aided Design Integr. Circuits Syst.*, vol. 27, no. 10, pp. 1725–1736, Oct. 2008.

[31] N. Patil et al., "Wafer-scale growth and transfer of aligned single-walled carbon nanotubes," *IEEE Trans. Nanotechnol.*, vol. 8, no. 4, pp. 498–504, Jul. 2009.

[32] N. Patil et al., "VMR: VLSI-compatible metallic carbon nanotube removal for imperfection-immune cascaded multi-stage digital logic circuits using carbon nanotube FETs," in *Proc. IEEE Int. Electron Devices Meeting (IEDM)*, Baltimore, MD, USA, 2009, pp. 1–4.

[33] N. Patil et al., "Scalable carbon nanotube computational and storage circuits immune to metallic and mispositioned carbon nanotubes," *IEEE Trans. Nanotechnol.*, vol. 10, no. 4, pp. 744–750, Jul. 2011.

[34] B. C. Paul et al., "Impact of a process variation on nanowire and nanotube device performance," *IEEE Trans. Electron Devices*, vol. 54, no. 9, pp. 2369–2376, Sep. 2007.

[35] A. Raychowdhury et al., "Variation tolerance in a multichannel carbon-nanotube transistor for high-speed digital circuits," *IEEE Trans. Electron Devices*, vol. 56, no. 3, pp. 383–392, Mar. 2009.

[36] J. A. Roy and I. L. Markov, "High-performance routing at the nanometer scale," *IEEE Trans. Comput.-Aided Design Integr. Circuits Syst.*, vol. 27, no. 6, pp. 1066–1077, Jun. 2008.

[37] R. Saito, G. Dresselhaus, and M. S. Dresselhaus, *Physical Properties of Carbon Nanotubes*. London, U.K.: Imperial College Press, 1998.

[38] N. Z. Shor, *Minimization Methods for Non-Differentiable Functions*. Berlin, Germany: Springer, 1985.

[39] M. M. Shulaker et al., "Carbon nanotube computer," *Nature*, vol. 501, no. 7468, pp. 526–530, 2013.

[40] M. M. Shulaker et al., "Sensor-to-digital interface built entirely with carbon nanotube FETs," *IEEE J. Solid-State Circuits*, vol. 49, no. 1, pp. 190–201, Jan. 2014.

[41] M. M. Shulaker et al., "Carbon nanotube circuit integration up to sub-20 nm channel length," *ACS Nano*, vol. 8, no. 4, pp. 3434–3443, 2014.

[42] M. M. Shulaker et al., "Monolithic 3D integration of logic and memory: Carbon nanotube FETs, resistive RAM, and silicon FETs," in *Proc. Int. Electron Devices Meeting (IEDM)*, San Francisco, CA, USA, 2014, pp. 27.4.1–27.4.4.

[43] M. M. Shulaker et al., "High-performance carbon nanotube field-effect transistors," in *Proc. Int. Electron Devices Meeting (IEDM)*, San Francisco, CA, USA, 2014, pp. 33.6.1–33.6.4.

[44] S. Sinha et al., "Design benchmarking to 7 nm with FinFET predictive technology models," in *Proc. ACM/IEEE Int. Symp. Low Power Electron. Design (ISLPED)*, Redondo Beach, CA, USA, 2012, pp. 15–20.

[45] (Apr. 2015). *Stanford University VSCNFET Model*. [Online]. Available: https://nano.stanford.edu/stanford-cnfet2-model

[46] L. Wei, D. J. Frank, L. Chang, and H.-S. P. Wong, "A non-iterative compact model for carbon nanotube FETs incorporating source exhaustion effects," in *Proc. Int. Electron Devices Meeting (IEDM)*, Baltimore, MD, USA, 2009, pp. 1–4.

[47] H. Wei et al., "Monolithic three-dimensional integration of carbon nanotube FET complementary logic circuits," in *Proc. Int. Electron Devices Meeting (IEDM)*, Washington, DC, USA, 2013, pp. 19.7.1–19.7.4.

[48] N. H. Weste and D. M. Harris, *CMOS VLSI Design*. Boston, MA, USA: Pearson/Addison Wesley, 2005.

[49] J. Zhang, N. Patil, A. Hazeghi, and S. Mitra, "Carbon nanotube circuits in the presence of carbon nanotube density variations," in *Proc. 46th ACM/IEEE Design Autom. Conf. (DAC)*, San Francisco, CA, USA, 2009, pp. 71–76.

[50] J. Zhang, N. Patil, and S. Mitra, "Probabilistic analysis and design of metallic-carbon-nanotube-tolerant digital logic circuits," *IEEE Trans. Comput.-Aided Design Integr. Circuits Syst.*, vol. 28, no. 9, pp. 1307–1320, Sep. 2009.

[51] J. Zhang et al., "Carbon nanotube correlation: Promising opportunity for CNFET circuit yield enhancement," in *Proc. 47th Design Autom. Conf. (DAC)*, Anaheim, CA, USA, 2010, pp. 889–892.

[52] J. Zhang, N. Patil, H.-S. P. Wong, and S. Mitra, "Overcoming carbon nanotube variations through co-optimized technology and circuit design," in *Proc. Int. Electron Devices Meeting (IEDM)*, Washington, DC, USA, 2011, pp. 4.6.1–4.6.4.

[53] J. Zhang, "Variation-aware design of carbon nanotube digital VLSI circuits," Ph.D. dissertation, Dept. Electr. Eng., Stanford Univ., Stanford, CA, USA, 2011.

[54] J. Zhang et al., "Carbon nanotube robust digital VLSI," *IEEE Trans. Comput.-Aided Design Integr. Circuits Syst.*, vol. 31, no. 4, pp. 453–471, Apr. 2012.

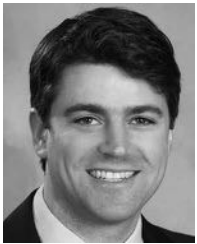

**Gage Hills** received the B.S. degree in electrical engineering and computer science from Yale University, New Haven, CT, USA, in 2007, and the M.S. degree in electrical engineering from Stanford University, Stanford, CA, USA, in 2012, where he is currently pursuing the Ph.D. degree.

His current research interests include experimental demonstrations and statistical modeling of very large-scale carbon nanotube-based digital circuits.

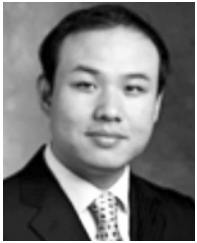

**Jie Zhang** (S'05–M'11) received the B.E. degree in electronic engineering from Tsinghua University, Beijing, China, in 2006, and the M.S. and Ph.D. degrees in electrical engineering from Stanford University, Stanford, CA, USA, in 2008 and 2011, respectively.

Since 2011, he has been with Google, Inc., Mountain View, CA, USA. His current research interests include modeling and simulation of carbon nanotube-based devices and circuits, with a focus on variation-aware design and optimization.

Dr. Zhang was a recipient of the Stanford Graduate Fellowship and the Top-Class Scholarship from Tsinghua University.

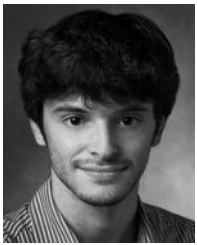

**Max Marcel Shulaker** received the B.S. in electrical engineering from Stanford University, Stanford, CA, USA, in 2011, where he is currently pursuing the Ph.D. degree, and researches on experimentally demonstrating nanosystems with emerging technologies.

His current research interests include realizing increased levels of integration for carbon nanotube-based digital logic circuits.

Mr. Shulaker is a Stanford Graduate Fellow and a Fannie and John Hertz Fellow.

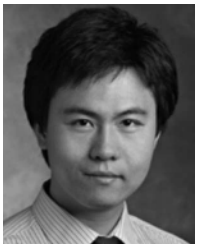

**Hai Wei** (S'09) received the B.S. (Hons.) degree in microelectronics from Tsinghua University, Beijing, China, in 2007, and the M.S. and Ph.D. degrees in electrical engineering from Stanford University, Stanford, CA, USA, in 2010 and 2014, respectively.

His current research interests include design and fabrication of carbon nanotube field-effect transistors and circuits and monolithic 3-D integrated circuits.

Mr. Wei was a recipient of the Stanford School of Engineering Fellowship Award.

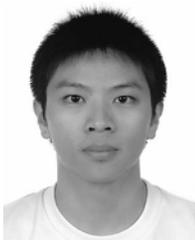

**Chi-Shuen Lee** received the B.S. degree in electrical engineering from National Taiwan University, Taipei, Taiwan, in 2011, and the M.S. degree in electrical engineering from Stanford University, Stanford, CA, USA, in 2014, where he is currently pursuing the Ph.D. degree.

He is currently involved in the compact modeling of carbon nanotube field-effect transistors and performance benchmarking of digital systems based on emerging CMOS technologies in the sub-10 nm regime. His current research interests include modeling and simulation of nanoscale transistors and performance benchmarking.

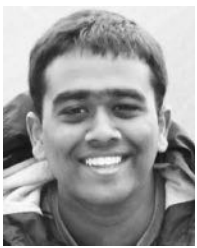

**Arjun Balasingam** received the Diploma degree from Archbishop Mitty High School, San Jose, CA, USA, in 2014, as the Valedictorian of his class. He is currently pursuing the B.S. degree with the School of Engineering, Stanford University, Stanford, CA, USA.

His current research interests include electrical engineering, mathematics, and computer science.

Mr. Balasingam was selected as a Siemens Competition Regional Finalist in 2012 and a Semifinalist in 2013 for his work on a variety of engineering research projects. He was selected as an Intel Science Talent Search Semifinalist in 2014. He was a recipient of the IEEE Silicon Valley Engineering Council Education Award in 2014.

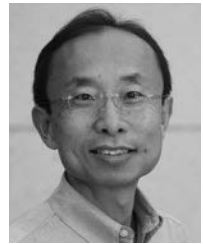

**H.-S. Philip Wong** (F'01) received the B.Sc. (Hons.) degree from the University of Hong Kong, Hong Kong, the M.S. degree from Stony Brook University, New York, NY, USA, and the Ph.D. degree from Lehigh University, Bethlehem, PA, USA.

From 1988 to 2004, he was with IBM T.J. Watson Research Center, Yorktown Heights, NY, USA, where he has been Research Staff Member, Manager, and Senior Manager. As Senior Manager, he was responsible for shaping and executing IBM's strategy on nanoscale science and technology as well as exploratory silicon devices and semiconductor technology. He joined Stanford University, Stanford, CA, USA, as a Professor of Electrical Engineering, in 2004, where he is the Willard R. and Inez Kerr Bell Professor with the School of Engineering. His academic appointments include the Chair of Excellence of the French Nanosciences Foundation, Grenoble, France, a Guest Professor with Peking University, Beijing, China, an Honorary Professor with the Institute of Microelectronics of Chinese Academy of Sciences, Beijing, a Visiting Chair Professor of Nanoelectronics with Hong Kong Polytechnic University, Hong Kong, and the Honorary Doctorate degree from the Institut Polytechnique de Grenoble, Grenoble. His current research interests include carbon electronics, 2-D layered materials, wireless implantable biosensors, directed self-assembly, nanoelectro mechanical relays, device modeling, brain-inspired computing, and nonvolatile memory devices such as phase change memory and metal oxide resistance change memory.

Prof. Wong has served as an Elected Member of the Electron Devices Society AdCom from 2001 to 2006, an Editor-in-Chief of the IEEE TRANSACTIONS ON NANOTECHNOLOGY from 2005 to 2006, the Sub-Committee Chair of the International Solid-State Circuit Conference from 2003 to 2004, the General Chair of the International Electron Devices Meeting in 2007. He is the IEEE ExCom Chair of the Symposia of Very Large-Scale Integration Technology and Circuits.

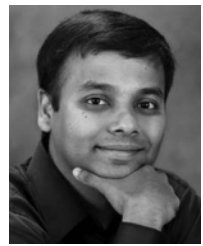

**Subhasish Mitra** (F'13) directs the Robust Systems Group in the Department of Electrical Engineering and the Department of Computer Science of Stanford University, where he is the Chambers Faculty Scholar of Engineering. Before joining Stanford, he was a Principal Engineer at Intel.

Prof. Mitra's research interests include robust systems, VLSI design, CAD, validation and test, emerging nanotechnologies, and emerging neuroscience applications. His X-Compact technique for test compression has been key to cost-effective manufacturing and high-quality testing of a vast majority of electronic systems, including numerous Intel products. X-Compact and its derivatives have been implemented in widely-used commercial Electronic Design Automation tools. His work on carbon nanotube imperfection-immune digital VLSI, jointly with his students and collaborators, resulted in the demonstration of the first carbon nanotube computer, and it was featured on the cover of NATURE. The NSF presented this work as a Research Highlight to the US Congress, and it also was highlighted as "an important, scientific breakthrough" by the BBC, Economist, EE Times, IEEE Spectrum, MIT Technology Review, National Public Radio, New York Times, Scientific American, Time, Wall Street Journal, Washington Post, and numerous other organizations worldwide.

Prof. Mitra's honors include the Presidential Early Career Award for Scientists and Engineers from the White House, the highest US honor for early-career outstanding scientists and engineers, ACM SIGDA/IEEE CEDA A. Richard Newton Technical Impact Award in Electronic Design Automation, "a test of time honor" for an outstanding technical contribution, and the Intel Achievement Award, Intel's highest corporate honor. He and his students published several award-winning papers at major venues: IEEE/ACM Design Automation Conference, IEEE International Solid-State Circuits Conference, IEEE International Test Conference, IEEE Transactions on CAD, IEEE VLSI Test Symposium, Intel Design and Test Technology Conference, and the Symposium on VLSI Technology. At Stanford, he has been honored several times by graduating seniors "for being important to them during their time at Stanford."

Prof. Mitra has served on numerous conference committees and journal editorial boards. He served on DARPA's Information Science and Technology Board as an invited member. He is a Fellow of the ACM and the IEEE.

# Appendix

TABLE VI
RESULTS OF CNFET DEVICE PARAMETER OPTIMIZATION (SECTION VII) FOR EACH NODE AND VALUE OF $V_{DD}$, USING THE VSCNFET MODEL [45].
"FIXED" VARIABLES ARE HELD CONSTANT DURING OPTIMIZATION; "OPTIMIZED" VARIABLES ARE SWEPT TO OPTIMIZE EDP (SECTION VII-A).
VALUES FOR STANDARD CELL PARASITICS ($R_{GCONT}$, $R_{WPT}$, $R_{SDCONT}$, $C_{IN\text{-}OUT}$, $C_{IN\text{-}RAILS}$, $C_{OUT\text{-}RAILS}$) ARE SHOWN FOR AN INVERTER WITH
ONE N-TYPE CNFET AND ONE P-TYPE CNFET, EACH WITH WIDTH EQUAL TO THE MAXIMUM FINGER WIDTH ($W_{FINGER,MAX}$);
STANDARD CELL CIRCUIT SCHEMATICS ARE ANNOTATED WITH THESE PARASITICS AS ILLUSTRATED IN [44].

| Parameter | Source | Technology parameter value | | | | | | |
|---|---|---|---|---|---|---|---|---|
| Technology node label | [18] | 14 nm | 10 nm | 7 nm | 5 nm | 5 nm | 5 nm | 5 nm |
| Year | [18] | 2014 | 2016 | 2018 | 2020 | 2020 | 2020 | 2020 |
| Supply voltage, $V_{DD}$ (V) | Fixed | 0.50 | 0.50 | 0.50 | 0.50 | 0.45 | 0.40 | 0.35 |
| CNFET contacted gate pitch, $L_{PITCH}$ (nm) | [18] | 64 | 56.6 | 45 | 35.8 | 35.8 | 35.8 | 35.8 |
| CNFET gate length, $L_G$ (nm) | Optimized | 9.8 | 9.5 | 8.9 | 8.4 | 8.8 | 9.3 | 9.7 |
| CNT-metal contact length, $L_C$ (nm) | Optimized | 44.3 | 38.1 | 29 | 21.5 | 20.9 | 20.3 | 19.8 |
| CNFET extension region length, $L_{EXT}$ (nm) | Optimized | 4.95 | 4.5 | 3.55 | 2.95 | 3.05 | 3.1 | 3.15 |
| CNFET nominal on-current, $I_{ON}$ (mA / μm) | Optimized | 1.914 | 1.826 | 1.602 | 1.348 | 1.066 | 0.800 | 0.552 |
| CNFET nominal off-current, $I_{OFF}$ (nA / μm) | [18] | 100 | 100 | 100 | 100 | 100 | 100 | 100 |
| CNFET nominal on/off ratio | Optimized | 19e3 | 18e3 | 16e3 | 13e3 | 11e3 | 8e3 | 5e3 |
| CNFET flat-band voltage, $V_{FB}$ (V) | Optimized | 0.007 | 0.010 | 0.018 | 0.026 | 0.018 | 0.010 | 0.005 |
| Nominal CNT-CNT spacing, $s$ (nm) | Fixed [46] | 4 | 4 | 4 | 4 | 4 | 4 | 4 |
| CNT diameter, $d_{CNT}$ (nm) | Fixed [31] | 1.3 | 1.3 | 1.3 | 1.3 | 1.3 | 1.3 | 1.3 |
| CNFET gate oxide thickness, $T_{OX}$ (nm) | [18] | 2.57 | 2.51 | 2.46 | 2.42 | 2.42 | 2.42 | 2.42 |
| CNFET gate oxide dielectric constant, $K_{OX}$ | [18] | 13 | 14 | 15 | 16 | 16 | 16 | 16 |
| CNFET gate equivalent oxide thickness, $EOT$ (nm) | [18] | 0.771 | 0.699 | 0.640 | 0.590 | 0.590 | 0.590 | 0.590 |
| CNFET gate height, $h_G$ (nm) | [18] | 20 | 20 | 15 | 15 | 15 | 15 | 15 |
| CNFET contact height, $h_C$ (nm) | [23] | 20 | 20 | 15 | 15 | 15 | 15 | 15 |
| Dielectric constant of spacer, $K_{SPA}$ | [18] | 2.775 | 2.59 | 2.59 | 2.31 | 2.31 | 2.31 | 2.31 |
| M1 wire resistivity (μΩ-cm) | [18] | 4.62 | 4.77 | 5.41 | 6.35 | 6.35 | 6.35 | 6.35 |
| M1 wire resistance (Ω / μm) | [18] | 23.7 | 29.8 | 53.4 | 99.1 | 99.1 | 99.1 | 99.1 |
| M1 wire capacitance (fF / μm) | [18] | 0.19 | 0.19 | 0.19 | 0.17 | 0.17 | 0.17 | 0.17 |
| M1 aspect ratio | [18] | 1.9 | 2.0 | 2.0 | 2.0 | 2.0 | 2.0 | 2.0 |
| Standard cell height (μm) | [18], [25] | 0.768 | 0.679 | 0.540 | 0.430 | 0.430 | 0.430 | 0.430 |
| Standard cell width (relative) | [18], [25] | 1.0 | 0.89 | 0.70 | 0.56 | 0.56 | 0.56 | 0.56 |
| Standard cell area (relative) | [18], [25] | 1.0 | 0.78 | 0.49 | 0.31 | 0.31 | 0.31 | 0.31 |
| Standard cell maximum finger width, $W_{FINGER,MAX}$ (μm) | [18], [25] | 0.280 | 0.240 | 0.200 | 0.160 | 0.160 | 0.160 | 0.160 |
| CNT sampling region width (nm) | Fixed | 20 | 20 | 20 | 20 | 20 | 20 | 20 |
| Minimum CNFET width in library (nm) | Fixed | 20 | 20 | 20 | 20 | 20 | 20 | 20 |
| Minimum CNFET width (1/2 $L_{PITCH}$) (nm) | Fixed | 32 | 28.8 | 22.5 | 17.6 | 17.6 | 17.6 | 17.6 |
| Standard cell input-to-gate wire resistance, $R_{GCONT}$ (Ω) | [18], [44] | 150 | 155 | 176 | 206 | 206 | 206 | 206 |
| Standard cell wire resistance per track, $R_{WPT}$ (Ω) | [18], [44] | 1.7 | 2.0 | 2.8 | 4.7 | 4.7 | 4.7 | 4.7 |
| Standard cell source/drain wire resistance, $R_{SDCONT}$ (Ω) | [18], [44] | 100 | 113 | 142 | 179 | 179 | 179 | 179 |
| Standard cell input-to-output capacitance, $C_{IN\text{-}OUT}$ (aF) | [18], [44] | 37.5 | 33.2 | 26.4 | 21.0 | 21.0 | 21.0 | 21.0 |
| Standard cell input-to-supply rail capacitance, $C_{IN\text{-}RAILS}$ (aF) | [18], [44] | 29.5 | 26.1 | 20.7 | 16.5 | 16.5 | 16.5 | 16.5 |
| Standard cell output-to-supply rail capacitance, $C_{OUT\text{-}RAILS}$ (aF) | [18], [44] | 38.0 | 33.6 | 26.7 | 21.3 | 21.3 | 21.3 | 21.3 |

## VI. CNT VARIATIONS & CNT CORRELATION

CNTs are subject to the following CNT-specific variations:
1) *CNT Type Variations*: CNTs can be either metallic (m-CNT) or semiconducting (s-CNT) [50].
2) *CNT Density Variations*: described in Section II.
3) *CNT Diameter Variations*: the diameter of a CNT is a function of its chirality, and can lead to changes in CNFET threshold voltage and on-current [34].
4) *CNT Alignment Variations*: mis-positioned CNTs cause random alignment angles with respect to the CNT growth direction, resulting in variations in CNFET channel length [30].
5) *CNT Doping Variations*: CNFETs require heavily doped source and drain extension regions to achieve small parasitic series resistance. Variations in the doping concentration lead to variation in series resistance [9].

*A. Improving $p_m$: Diminishing Returns*

As described in Section II-D, reducing $p_m$ past 1% suffers from diminishing returns and can be insufficient to meet design goals [16], [52]; Fig. 16 illustrates that $p_m = 0.1\%$ does not achieve delay penalty ≤5% for the OpenSPARC modules at the 5 nm node.

*B. Gaussian Approximation of CNT Count Distributions*

To validate the Gaussian approximation to the CNT count distribution (as described in Section III-A), we sample the circuit delay cumulative distribution function (CDF) via MC SSTA (Section III-A) for each of two cases: using discrete (non-negative integer) CNT count variables [49], and using the Gaussian approximation. For example, in the case of the 5 nm "gkt" OpenSPARC module, the Gaussian approximation underestimates the median delay (where the CDF is equal to 50%) by only 0.07%, and overestimates the delay spread (measured as the width between the points where the CDF is equal to 5% and 95%) by only 0.8% (Fig. 16). We conclude that the Gaussian approximation is sufficient for our exploration purposes.

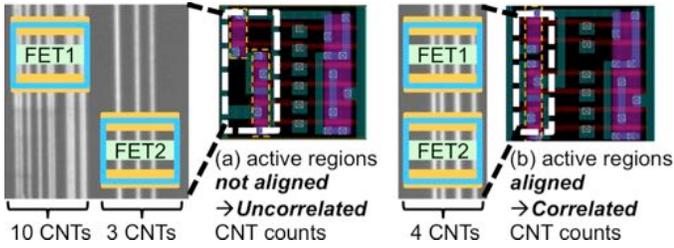

Fig. 17. Aligned-active layout (Section II-A). AOI222_X1 standard cell [25] before (a) and after (b) active alignment.

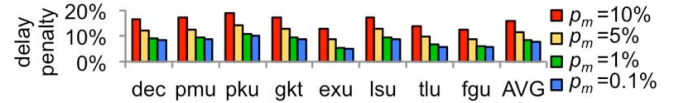

Fig. 15. Delay penalty improvement due to improving $p_m$ (5 nm node, OpenSPARC modules, after steps 1-3 in Fig. 3). $IDC = 0.50$, $p_{Rs} = 4\%$, $p_{Rm} = 99.99\%$. Improving $p_m$ from 10% to 0.1% improves count-limited yield from ≥99.98% to ≥99.995%. However, despite $p_m = 0.1\%$, delay penalty can be >10%; thus, improving $p_m$ alone can be insufficient to meet design goals.

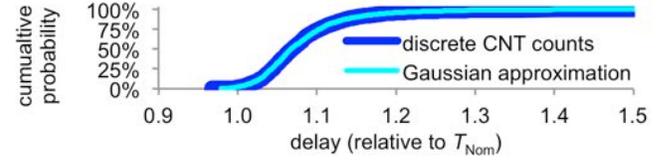

Fig. 16. Estimated cumulative distribution function (CDF) of maximum path delay (5 nm "gkt" OpenSPARC module: 2000 MC trials).

## VII. CNFET DEVICE- & CIRCUIT-LEVEL MODELING

To efficiently evaluate delay penalties and *PNMV* for the OpenSPARC modules, we use variation-aware logic gate timing, energy, and SNM models that are built using a SPICE-compatible CNFET compact device model [23], which is based on the virtual source model [21]. This virtual source CNFET (VSCNFET) device model accounts for several non-idealities including (but not limited to) direct source-to-drain tunneling leakage current, parasitic gate-to-plug capacitance, fringing capacitance, source/drain extension region resistance, and CNT-metal contact resistance [45]. It has been calibrated with experimental data from 15 nm gate length CNFETs [23] and with data from NEGF-based (non-equilibrium Green's function) simulations of 5 nm gate length CNFETs [4].

We leverage the VSCNFET model to extract timing, energy, and noise margin information from SPICE simulations to build variation-aware logic gate timing, energy, and SNM models for each standard cell in our standard cell library, which is derived from the Nangate 15 nm Open Cell Library (OCL) [25]. We use the standard cell height, width, area, and maximum finger width from the Nangate 15 nm OCL for our 14 nm standard cell library, and then scale each of these dimensions by the ratio of the contacted gate pitch ($L_{PITCH}$) given by the 2013 edition of the International Technology Roadmap for Semiconductors (ITRS) [18]. $L_{PITCH}$ (as well as other device- and circuit-level parameters, including oxide thickness and wire resistivity) for the 14, 10, 7, and 5 nm technology nodes is taken according to the "Node Range" Labeling in the Process Integration, Devices, and Structures (PIDS) table for High Performance Logic Technology Requirements. All of our variation-aware models account for standard cell parasitics, including wire resistance to the source, drain, and gate of each CNFET, wire track resistance, and capacitance between the input, output, and supply rails (using experimentally measured values from [44] for the 14 nm node, which are then scaled for other nodes using parameters from the ITRS) [44]. See Table VI for results.

### A. CNFET Device Parameter Optimization

Before building our variation-aware models for each standard cell, we first optimize the CNFET device parameters (i.e., parameters that define the geometry of the CNFETs and affect their electrical characteristics, details below) to target a high performance CNFET technology (as opposed to a low power CNFET technology; e.g., high performance versus low power options for standard cell libraries in silicon-CMOS circuits are often distinguished by the transistor threshold voltage and off-state leakage current) [18]. For example, these CNFET device parameters include the CNFET channel length, CNT-metal contact length, and the CNT diameter, all of which affect CNFET electrical characteristics (e.g., threshold voltage, parasitic capacitances, on-current, off-current, sub-threshold slope, etc.). We choose to optimize the CNFET device parameters so as to minimize the EDP of an inverter with fan-out (FO) equal to four (i.e., the output load capacitance is four times as large as the input gate capacitance): a common metric for performance benchmarking and technology assessment [9]. We refer to this metric as $EDP_{FO4}$, where $EDP_{FO4} = E_{FO4}T_{FO4}$, $T_{FO4}$ is the average of the rise delay (falling input/rising output) and the fall delay (rising input/falling output), and $E_{FO4}$ is the average switching energy per transition. We perform the following CNFET device parameter optimization to minimize $EDP_{FO4}$ (for each node and for each value of $V_{DD}$): $L_{PITCH}$ is held constant, and then $L_G$ (gate length), $L_C$ (CNT-metal contact length), $L_{EXT}$ (CNT extension length, which refers to the un-gated region of the CNT between the gate and the source/drain contact, Fig. 1 [23]), and $V_{FB}$ (flat-band voltage, which offsets the threshold voltage), are swept using the VSCNFET model to minimize $EDP_{FO4}$, subject to the constraint that the CNFET off-current $I_{OFF} \leq 100$ nA/μm in the nominal case ($I_{OFF} = 100$ nA/μm is the target for high performance logic specified by the ITRS) [18]. Note that, $L_{PITCH} = L_C + L_G + 2L_{EXT}$ [23]. For each combination of CNFET device parameters, we simulate $EDP_{FO4}$ using SPICE and the VSCNFET model and then select the CNFET device parameters that minimize $EDP_{FO4}$. CNFET device parameter optimization results (as well as additional parameters, including gate dielectric constant, contact height, CNT-CNT spacing, etc.) are provided in Table VI. In particular, we illustrate the optimized $L_C$ values (Table VI) in Fig. 18, demonstrating that $EDP_{FO4}$ is highly sensitive to CNT-metal contact resistance ($R_C$).

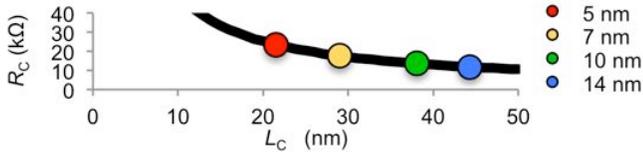

Fig. 18. Contact resistance ($R_C$) versus contact length ($L_C$) per CNT for the VSCNFET model (diameter-independent case) [23]. CNFET device parameter optimization results are shown for each node. $R_C$ versus $L_C$ uses a transmission line model that is calibrated using experimental values extracted from CNFETs with $L_C$ ranging from 20-300 nm [23].

The optimized CNFET device parameters are then used as inputs to the VSCNFET model for all SPICE simulations to analyze $I_{ON}$ variations (Fig. 2) and to build variation-aware logic gate timing, energy, and SNM models to evaluate circuit-level performance metrics (e.g., delay penalty, *PNMV*, and $\Delta E$) for each node and for multiple values of $V_{DD}$ ($V_{DD} = 0.50$ V to compare technology nodes and $V_{DD}$ is swept down to 0.35 V in 0.05 V increments at the 5 nm node to evaluate the impact of $V_{DD}$ scaling).

*B. Physical Circuit Design for Circuit-Level Analysis*

For circuit-level analysis, the OpenSPARC modules are synthesized using Synopsys Design Compiler (targeting the nominal case), Capo [36] is used for placement, and FLUTE [8] is used to estimate wire lengths, with wire parasitics computed using parameters from the ITRS [18]. The full design and analysis flow is shown in Fig. 3 (Section II).

*C. Selective Upsizing*

We leverage the following selective transistor/logic gate upsizing algorithm (i.e., selective upsizing, inspired by [52]) to minimize circuit EDP (Section II-A) and to reduce circuit delay penalty (Section II-B). We first sort all standard cells according to their fan-out (fan-out: the ratio of the output load capacitance to the minimum input capacitance on any input) in the nominal case. Next, we upsize the standard cell with the largest fan-out by incrementing its drive strength (e.g., INV_X1 becomes INV_X2) and then re-sort all of the standard cells according to their fan-out; a parameterized number $k_{SelUpsize} \geq 0$ of the standard cells are upsized sequentially in this manner (note that, each standard cell can potentially be upsized multiple times). If a standard cell cannot be upsized because it is at its maximum drive strength (e.g., INV_X64 is the strongest inverter in our library), then the standard cell with the largest fan-out that can be upsized (i.e., it is not at its maximum drive strength) is upsized instead.

VIII. VARIATION-AWARE TIMING/ENERGY MODEL

*A. Timing Model Generation*

The VSCNFET SPICE model and CNFET device parameter optimization results (Table VI) are used to build a variation-aware logic gate timing model that computes CNFET logic gate delay and output slew as functions of the CNT count in each sampling region (see Fig. 8 in Section III-B for details on the CNT count variables).

To build this model, we perform over 2000 SPICE simulations for each input pin of each logic stage in our standard cell library, varying the input slew rate ($t_{InSlew}$), the load capacitance ($C_{Load}$), and the CNT counts ($n_1, n_2, ...$) of each sampling region. Four values of $C_{Load}$ and six values of $t_{InSlew}$ are analyzed for each logic stage to emulate typical operating conditions in a digital system ($C_{Load}$: fan-out = 2, 4, 6, and 8; $t_{InSlew}$: 1-64 ps). For each of the 24 combinations of ($C_{Load}$, $t_{InSlew}$), we sample 100 random values of CNT count from the CNT count distribution (using processing parameters defined in Table I), for a total of 2400 simulations for each input pin of each logic stage. We then calibrate timing models $f(t_{InSlew}, C_{Load}, n_1, n_2, ...)$ to the delay ($d$) and to the output slew ($t_{OutSlew}$) values extracted from the SPICE simulations.

*B. Timing Model Linearization*

Solving for $d$ in (41) (using the nonlinear timing model in [53]) is not trivial, and may involve a numerical method that requires significantly more computational effort than a model of the form $d = CV/i$ [48]:

$$d = (C_{Par} + C_{Load})V_{DD}/(i_1 \min(2d/t_{InSlew}, 1) + i_2). \quad (41)$$

Additionally, $t_{InSlew}$ must be determined for each input pin of each logic stage and must propagate through the circuit (as it affects the delay of subsequent logic stages), which further increases the computation time. To obtain the linearized model in (42), we linearize (41) with the following procedure:

$$d = (C_{Par} + C_{Load})V_{DD}/i_{Drive}. \quad (42)$$

1) Perform static timing analysis with the nonlinear timing model to calculate $T_{Nom}$ (Section II). This also yields the nominal delays and input slew rates for each logic stage: $d_{Nom}$ and $t_{InSlewNom}$
2) Define a new parameter, $i_{Drive}$, which is an affine function of $i_1$ and $i_2$, and is therefore also an affine function of the region CNT counts:

$$i_{Drive} = i_1 \min(2d_{Nom}/t_{InSlewNom}, 1) + i_2. \quad (43)$$

3) Replace the denominator in (41) with the value of $i_{Drive}$ to create a first-order delay model of the form $d = CV/i$ [as in (42)] that gives the same value of $d$ as the nonlinear model in the nominal case. We choose to linearize the timing model around the nominal case so that it is independent of the processing parameter values. This enables delay factorization (Section III-A) to further improve computational efficiency.

*C. Timing Model Validation*

We validate our timing models using circuit modules synthesized from the ISCAS-85 benchmarks [15]. For each circuit module, we compare the critical path delay (in the nominal case) computed using our timing model versus SPICE simulations (using the VSCNFET model), according to the following three-step procedure.

1) For the EDP-optimal nominal design point [defined in Section II-A: ($E_{NomOpt}$, $T_{NomOpt}$) (1)], compute the nominal critical path delay using the timing model and record an arbitrary critical path.
2) Create a SPICE netlist of the cascaded standard cells on the recorded critical path (using the standard cell library described in Section VII) and instantiate capacitors to account for the capacitances of branches off the recorded critical path.

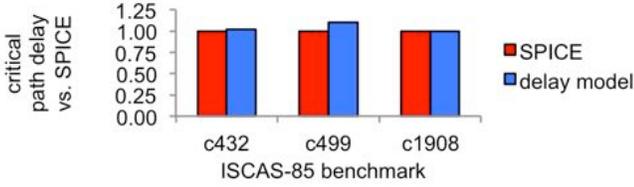

Fig. 19. Critical path delay comparison: timing model versus delay extracted from SPICE transient analysis, shown for three circuit modules synthesized from the ISCAS-85 benchmarks (5 nm node, after steps 1-3 in Fig. 3) [15].

TABLE VII
PROCESSING PARAMETER VALUES USED TO VALIDATE THE LINEARIZED TIMING MODEL (SECTION IV-A). $p_{Rm}$ = 99.99% FOR ALL TESTS

| Test | 1 | 2 | 3 | 4 | 5 | 6 | 7 | 8 |
|---|---|---|---|---|---|---|---|---|
| IDC | 0.50 | 0.50 | 0.50 | 0.50 | 0.35 | 0.10 | 0.25 | 0.25 |
| $p_m$ | 10% | 5% | 1% | 0.1% | 5% | 5% | 5% | 5% |
| $p_{Rs}$ | 5% | 5% | 5% | 5% | 5% | 5% | 5% | 2.5% |

3) Perform transient analysis in SPICE and extract the delay (as the time taken for the output to reach $V_{DD}/2$ from the time that the input reaches $V_{DD}/2$); compare it to the critical path delay computed using the timing model in step 1.

Results for three of the ISCAS-85 benchmarks are shown in Fig. 19; the timing model overestimates the SPICE-extracted delay by an average of 3.8%, which we conclude is sufficient for our exploration purposes.

In Section IV-A, the linearized timing model is validated against the nonlinear timing model using the processing parameter values in Table VII (see Section IV-A for details).

## IX. VARIATION-AWARE SNM MODEL & *PNMV*

### A. Minimum-Width Upsizing to Improve PNMV

As shown in Fig. 7(c) in Section II-C, additional minimum-width upsizing (in addition to minimum-width upsizing for count-limited yield, i.e., in Fig. 3 in Section II-A) can improve *PNMV* at the cost of energy due to statistical averaging. In some cases, however, target *PNMV* constraints (e.g., *PNMV* ≤ 0.001%) cannot be satisfied even for arbitrarily high energy cost (e.g., for an arbitrarily large minimum width, $W_{MIN}$, for minimum-width upsizing): once the minimum CNFET width exceeds the maximum width of a single CNFET in the standard cell library (limited by the standard cell height, see Table VI) [25], then multiple CNFETs connected in parallel (i.e., multiple fingers) are required to increase the effective CNFET width. For highly aligned CNTs with aligned-active layouts, adding fingers to increase the effective CNFET width does not achieve the full benefits of statistical averaging because of CNT correlation. For example, a CNFET consisting of two CNFET fingers that have perfectly correlated CNT counts exhibits the same magnitude of CNT count variations as either of the CNFET fingers alone (measured by the standard deviation of the CNT count relative to its mean), despite having twice the effective width.

### B. Variation-Aware SNM Model Construction

As described in Section III-B1: for each logic stage input in our standard cell library, we model the VTC parameters for every case in which that input is sensitized (considering all possible combinations of the other inputs). The VTC parameters are functions of the CNT counts of the p- and n-type CNFETs which: 1) are gated by that input and 2) connect the logic stage output to either $V_{DD}$ or ground through a series of CNFETs in the "on" state. As an example, consider the pull-down network of a 3-input CMOS logic stage (inputs: inA, inB, inC) which consists of 4 CNFETs: *A*1, *A*2, *B*, and *C*. *A*1 and *A*2 are both gated by inA, *B* is gated by inB, and *C* is gated by inC. *A*1 and *B* are connected in series between the logic stage output and ground, and so are *A*2 and *C* (forming 2 parallel paths each with 2 CNFETs connected in series). Assume that we are obtaining the VTC of the logic stage for inA. Then when the state of (inB, inC) is (1, 0), *A*1 connects the logic stage output to ground through a series of CNFETs in the "on" state (since *B* is "on"), and *A*2 does not (since *C* is "off"). When the state of (inB, inC) is (1, 1) (i.e., another state that sensitizes inA), then both *A*1 and *A*2 connect the logic stage output to ground through a series of CNFETs in the "on" state (since both *B* and *C* are "on"), etc.

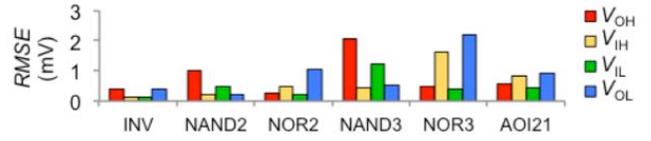

Fig. 20. SNM model calibration results for a subset of minimum-sized cells (e.g., NAND2_X1 for a NAND2 cell, for an arbitrary input) in our standard cell library (5 nm node, $V_{DD}$ = 0.35 V). *RMSE* ≤ 2.5 mV in all cases.

The VTC parameters are modeled separately for each input of a logic stage. For example, NAND2 gate *U*4 in Fig. 8 (Section III-B) consists of 4 CNFETs: 2 are gated by in1 (*P*4,1 and *N*4,1) and 2 are gated by in2 (*P*4,2 and *N*4,2). The VTC parameters for in1 are functions of $n_P^{(P4,1)}$ and $n_N^{(N4,1)}$ (i.e., the CNT counts of CNFETs *P*4,1 and *N*4,1), and the VTC parameters for in2 are functions of $n_P^{(P4,2)}$ and $n_N^{(N4,2)}$ (i.e., the CNT counts of CNFETs *P*4,2 and *N*4,2). Hence, two instances of *T* are required in the variation-aware SNM model (14): $T^{(NAND2\_X1\text{-}in1)}$ and $T^{(NAND2\_X1\text{-}in2)}$.

### C. SNM Model Calibration

Section III-B1 describes how we build the variation-aware SNM model for each input of each logic stage in our standard cell library. The error of this model versus SPICE simulations is shown in Fig. 20; for each VTC parameter, the root-mean-square error [*RMSE*, defined in (44), e.g., for $V_{OH}$] quantifies the error in the SNM model (14) versus VTC parameters extracted from 2000 SPICE simulations [e.g., $V_{OH,i}^{(VTC)}$ is the value of $V_{OH}$ extracted from the VTC for the *i*th SPICE simulation (of *q* = 2000 simulations), and $V_{OH,i}^{(model)}$ is its modeled value (14)].

$$RMSE = \sqrt{\frac{1}{q}\sum_{i=1}^{q}\left(V_{OH,i}^{(VTC)} - V_{OH,i}^{(model)}\right)^2} \quad (44)$$

### D. Rapid Analysis of Circuit PNMV

As described in Section III-B2, the SNM constraints on the CNFET CNT count variables are expressed using a matrix $H \in \mathbb{R}^{c \times t}$, such that satisfying $Hs \leqslant \mathbf{0}$ [(22) in Section III-B2] is equivalent to satisfying all SNM constraints in the circuit.

As an example, (45) shows the constraints $Hs \leqslant 0$ for the circuit in Fig. 8 [for gate pairs ($U1$, $U3$), ($U3$, $U5$), and ($U3$, $U6$)].[4] Each row in $H$ represents a single SNM constraint; e.g., the first row is an SNMH constraint for the gate pair ($U1$, $U3$) on the CNFET CNT count variables $n_P^{(P3,1)}$ and $n_N^{(N3,1)}$: the first two elements in vector $s$. In (45), each element of $H$ is subscripted by its row and column indices (e.g., $H_{1,2}$ for the 1st row, 2nd column of $H$). These terms are determined using equations similar to (20)-(21) along with all instances of $T$ in the variation-aware SNM model (14). For example, $H_{1,2}$ and $H_{2,1}$ are associated with the gate pair ($U1$, $U3$), hence they are computed using $T^{(\text{D-latch,Out})}$ for the output stage of D-latch $U1$ and $T^{(\text{INV\_X1})}$ for inverter $U3$.

$$\begin{array}{c}\text{SNMH}(U1,U3):\\ \text{SNML}(U1,U3):\\ \text{SNMH}(U3,U5):\\ \text{SNML}(U3,U5):\\ \text{SNMH}(U3,U6):\\ \text{SNML}(U3,U6):\\ \vdots\end{array}\begin{bmatrix}1 & H_{1,2} & 0 & 0 & 0 & 0 & \cdots\\ H_{2,1} & 1 & 0 & 0 & 0 & 0 & \cdots\\ 0 & 0 & 1 & H_{3,4} & 0 & 0 & \cdots\\ 0 & 0 & H_{4,3} & 1 & 0 & 0 & \cdots\\ 0 & 0 & 0 & 0 & 1 & H_{5,6} & \cdots\\ 0 & 0 & 0 & 0 & H_{6,5} & 1 & \cdots\\ \vdots & \vdots & \vdots & \vdots & \vdots & \vdots & \ddots\end{bmatrix}\begin{bmatrix}n_P(P3,1)\\ n_N(N3,1)\\ n_P(P5,1)\\ n_N(N5,1)\\ n_P(P6,1)\\ n_N(N6,1)\\ \vdots\end{bmatrix}\leqslant\begin{bmatrix}0\\0\\0\\0\\0\\0\\ \vdots\end{bmatrix} \quad (45)$$

Also described in Section III-B2, the relationship between the CNFET CNT count variables [$s \in \mathbb{R}^t$ in (23), for $t$ total CNFETs] and the sampling region CNT count variables [$n \in \mathbb{R}^r$ in (23), for $r$ total sampling regions] is expressed using a linear transformation $s = Bn$ (23), where $B \in \{0,1\}^{t \times r}$ has $B_{i,j} = 1$ if CNFET $i$ overlaps sampling region $j$, and $B_{i,j} = 0$ otherwise. For example, six rows of $B$ for the circuit in Fig. 8 are shown in (46). The first row represents the transformation from the sampling region CNT counts to the CNFET CNT count $n_P^{(P3,1)}$; since CNFET $P3,1$ overlaps sampling regions 1, 2, and 3, then $B_{1,1} = 1$, $B_{1,2} = 1$, $B_{1,3} = 1$ (all other values of $B$ in this row are 0), i.e., $n_P^{(P3,1)} = n_1 + n_2 + n_3$.

$$\begin{bmatrix}n_P(P3,1)\\ n_N(N3,1)\\ n_P(P5,1)\\ n_N(N5,1)\\ n_P(P6,1)\\ n_N(N6,1)\\ \vdots\end{bmatrix}=\begin{bmatrix}1&1&1&0&0&0&0&0&0&0&0&0&\cdots\\0&0&0&1&1&1&0&0&0&0&0&0&\cdots\\1&1&1&0&0&0&0&0&0&0&0&0&\cdots\\0&0&0&1&1&1&0&0&0&0&0&0&\cdots\\0&0&0&0&0&0&1&1&1&0&0&0&\cdots\\0&0&0&0&0&0&0&0&0&1&1&1&\cdots\\ \vdots & \vdots & \vdots & \vdots & \vdots & \vdots & \vdots & \vdots & \vdots & \vdots & \vdots & \vdots & \ddots\end{bmatrix}\begin{bmatrix}n_1\\n_2\\n_3\\ \vdots\\n_{10}\\n_{11}\\n_{12}\\ \vdots\end{bmatrix} \quad (46)$$

By substituting (46) into (45) and computing $K = HB$ (24), all SNM constraints [on the sampling region CNT count variables, i.e., $Kn \leqslant 0$ (25)] for the example circuit in Fig. 8 are expressed in (47). Although (47) can be used to determine *PNMV* (e.g., using an MC-based approach, as described in Section III-B2), it contains many noncritical SNM constraints, which can be eliminated to more efficiently compute *PNMV*.

$$\begin{bmatrix}1&1&1&H_{1,2}&H_{1,2}&H_{1,2}&0&0&0&0&0&0&\cdots\\ H_{2,1}&H_{2,1}&H_{2,1}&1&1&1&0&0&0&0&0&0&\cdots\\1&1&1&H_{3,4}&H_{3,4}&H_{3,4}&0&0&0&0&0&0&\cdots\\ H_{4,3}&H_{4,3}&H_{4,3}&1&1&1&0&0&0&0&0&0&\cdots\\0&0&0&0&0&0&1&1&1&H_{5,6}&H_{5,6}&H_{5,6}&\cdots\\0&0&0&0&0&0&H_{6,5}&H_{6,5}&H_{6,5}&1&1&1&\cdots\\ \vdots & \vdots & \vdots & \vdots & \vdots & \vdots & \vdots & \vdots & \vdots & \vdots & \vdots & \vdots & \ddots\end{bmatrix}\begin{bmatrix}n_1\\n_2\\n_3\\ \vdots\\n_{10}\\n_{11}\\n_{12}\\ \vdots\end{bmatrix}\leqslant\begin{bmatrix}0\\0\\0\\0\\0\\0\\ \vdots\end{bmatrix} \quad (47)$$

---

[4] Note that, a single logic gate can be a driving logic gate in multiple gate pairs. In this case, $U3$ drives both $U5$ and $U6$, so it is a driving logic gate in gate pairs ($U3$, $U5$), and ($U3$, $U6$); there are SNMH and SNML constraints for each of these gate pairs.

TABLE VIII
NUMBER OF SNM CONSTRAINTS AND CRITICAL SNM CONSTRAINTS FOR THE OPENSPARC MODULES (USING THE SYNTHESIZED NETLIST FOR EACH MODULE FOR THE 5 NM NODE WITH $V_{DD} = 0.50$ V)

| Open-SPARC module | Rows of standard cells | Width (µm) | Height (µm) | SNM constraints | Critical SNM constraints | Critical SNM constraints (% of total) |
|---|---|---|---|---|---|---|
| dec | 58 | 22.9 | 22.8 | 11.8K | 356 | 3.3% |
| pmu | 94 | 37.4 | 37.2 | 29.8K | 636 | 2.1% |
| pku | 98 | 38.9 | 38.8 | 32.2K | 754 | 2.3% |
| gkt | 100 | 40.2 | 39.6 | 26.6K | 536 | 2.0% |
| exu | 122 | 48.4 | 48.4 | 65.0K | 1226 | 1.9% |
| lsu | 207 | 82.8 | 82.4 | 141.0K | 1882 | 1.3% |
| tlu | 243 | 96.6 | 96.8 | 218.2K | 2684 | 1.2% |
| fgu | 275 | 110 | 110 | 364.6K | 3058 | 0.8% |

### E. Eliminating Noncritical SNM Constraints for PNMV

As described in Section III-B2, identifying and eliminating noncritical SNM constraints is crucial to efficiently determine *PNMV*; Table VIII illustrates that they can account for ≥99% of the total number of SNM constraints. Here, we describe how to systematically identify and eliminate noncritical SNM constraints. As an example, consider gate pairs ($U1$, $U3$) and ($U3$, $U5$) in Fig. 8 with the following SNMH constraints (15).

$$\text{SNMH}(U1, U3): V_{\text{OH}}^{(U1)} - V_{\text{IH}}^{(U3)} \geq SNM_R \quad (48)$$
$$\text{SNMH}(U3, U5): V_{\text{OH}}^{(U3)} - V_{\text{IH}}^{(U5)} \geq SNM_R. \quad (49)$$

We now describe a case where SNMH($U3$, $U5$) is always larger than SNMH($U1$, $U3$), meaning that (49) is a noncritical SNM constraint [i.e., it can only be violated if (48) is also violated].[5] First, consider the $V_{\text{IH}}$ terms in (48)-(49). We replace $V_{\text{IH}}^{(U3)}$ and $V_{\text{IH}}^{(U5)}$ with the variation-aware SNM model (14), using $T^{(\text{INV\_X1})}$ for both $V_{\text{IH}}$ terms (since both $U3$ and $U5$ are inverters):

$$V_{\text{IH}}^{(U3)} = V_{\text{IH0}}^{(\text{INV\_X1})} + V_{\text{IH1}}^{(\text{INV\_X1})} \log_{10}(n_P^{(P3,1)}/n_N^{(N3,1)}) \quad (50)$$
$$V_{\text{IH}}^{(U5)} = V_{\text{IH0}}^{(\text{INV\_X1})} + V_{\text{IH1}}^{(\text{INV\_X1})} \log_{10}(n_P^{(P5,1)}/n_N^{(N5,1)}). \quad (51)$$

These $V_{\text{IH}}$ terms are equivalent, since $n_P^{(P3,1)} = n_P^{(P5,1)}$ and $n_N^{(N3,1)} = n_N^{(N5,1)}$ (as the CNFETs in inverters $U3$ and $U5$ overlap the exact same sampling regions: Fig. 8). Thus, $V_{\text{IH}}^{(U3)} = V_{\text{IH}}^{(U5)}$, so the difference between (48) and (49) comes from the $V_{\text{OH}}$ terms. Using the variation-aware SNM model (14) [with $T^{(\text{D-latch,Out})}$ for the output stage of the D-latch], $V_{\text{OH}}^{(U1)} = T_{\text{VOH0}}^{(\text{D-latch,Out})}$ and $V_{\text{OH}}^{(U3)} = T_{\text{VOH0}}^{(\text{INV\_X1})}$. Therefore, if $T_{\text{VOH0}}^{(\text{INV\_X1})} > T_{\text{VOH0}}^{(\text{D-latch,Out})}$ (i.e., $V_{\text{OH}}$ for the inverter is higher than $V_{\text{OH}}$ for the output stage of the D-latch), then SNMH($U3$, $U5$) > SNMH($U1$, $U3$), signifying that (49) is a noncritical SNM constraint.

In the general case, noncritical SNM constraints can be identified and eliminated via the following procedure.
1) *SNM Constraint Partitioning*: each row of $K$ (25) (Section III-B2) corresponds to an SNM constraint on the sampling region CNT count variables. Partition the rows of $K$ into groups such that each row in a group constrains the *exact same* sampling region CNT count variables; i.e., given rows *i* and *j* in *K*, rows *i* and *j* should be partitioned into the same group if and only if: for all columns $1 \le k \le r$ (*K* has *r* columns since there are *r* sampling regions), either $K_{i,k} \neq 0$ and $K_{j,k} \neq 0$, or $K_{i,k} = 0$ and $K_{j,k} = 0$. For example, in (47), the first six rows of *K* are partitioned into two groups (labels are the row indices of *K*): {1, 2, 3, 4} and {5, 6}.

---

[5] If SNMH($U3$, $U5$) < $SNM_R$ [i.e., (49) is violated], then SNMH($U1$, $U3$) < SNMH($U3$, $U5$) implies SNMH($U1$, $U3$) < $SNM_R$ [i.e., (48) is also violated].

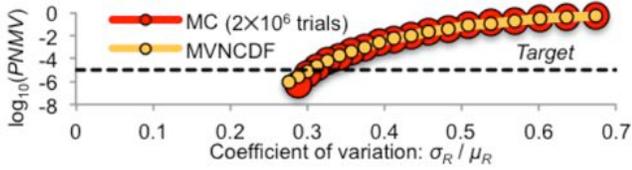

Fig. 21. Validation of MVNCDF-based *PNMV* calculation (31), shown for the 5 nm "dec" OpenSPARC module. The coefficient of variation is the standard deviation of the sampling region CNT count relative to the mean: it depends on the processing parameter values, and we sweep it by changing *IDC* with $p_m = 1\%$, $p_{Rs} = 4\%$, $p_{Rm} = 99.99\%$. $2 \times 10^6$ MC trials are required to reject the hypothesis that $PNMV > 0.001\%$ with 90% statistical power. The root-mean-squared percentage error is ≤10% (for *PNMV* values computed with statistical power ≥90%).

TABLE IX
VARIABLES IN THE FULL-CIRCUIT *PNMV* FORMULATION

| Variable | ∈ | Description |
|---|---|---|
| $t$ | $\mathbb{Z}^+$ | Total number of CNFET CNT count variables |
| $c$ | $\mathbb{Z}^+$ | Total number of SNM constraints |
| $s$ | $\mathbb{R}^t$ | Vector of CNFET CNT counts |
| $H$ | $\mathbb{R}^{c \times t}$ | Constraints on the CNFET CNT counts: $Hs \preccurlyeq 0$ |
| $r$ | $\mathbb{Z}^+$ | Total number of sampling regions |
| $\mu_R$ | $\mathbb{R}$ | Mean of the sampling region CNT count distribution |
| $\sigma_R$ | $\mathbb{R}$ | Standard deviation of the sampling region CNT count distribution |
| $n$ | $\mathbb{R}^r$ | Vector of sampling region CNT counts: normally distributed with mean $\mu_R$ & standard deviation $\sigma_R$ |
| $x$ | $\mathbb{R}^r$ | Vector of unit Gaussian random variables |
| $B$ | $\mathbb{Z}^{t \times r}$ | Transformation from the sampling region CNT counts to the CNFET CNT counts: $s = Bn$ |
| $K$ | $\mathbb{R}^{c \times r}$ | Constraints on the sampling region CNT counts: $K = HB$, $Kn \preccurlyeq 0$ |
| $p$ | $\mathbb{Z}^+$ | Total number of critical SNM constraints |
| $\widetilde{K}$ | $\mathbb{R}^{p \times r}$ | Constraints on the sampling region CNT counts, after removing noncritical SNM constraints: $\widetilde{K}n \preccurlyeq 0$ |
| $C$ | $\mathbb{R}^{p \times p}$ | Covariance matrix for MVNCDF formulation: $C = \widetilde{K}\widetilde{K}^T$ |
| $b$ | $\mathbb{R}^p$ | Upper bound for MVNCDF formulation: $b = -\widetilde{K}\mathbf{1}$ |
| $y$ | $\mathbb{Z}^+$ | Number of rows of standard cells |

2) *Noncritical SNM Constraint Identification*: compare each row *i* in *K* against each other row *j* in *K* that is partitioned into the same group, using the following comparison criterion: if $K_{i,k} \le K_{j,k}$ for all columns $1 \le k \le r$, then the constraint in row *i* is noncritical, since it cannot be violated without simultaneously violating the constraint in row *j*. This property holds for $s \succ 0$ (i.e., the vector of CNFET CNT counts is strictly greater than 0, meaning that there is no CNT count failure; CNT count failure and count-limited yield are accounted for separately using minimum-width upsizing for count-limited yield: Section II-A).

3) *Noncritical SNM Constraint Elimination*: remove all rows of *K* that correspond to the noncritical SNM constraints identified in step 2.

As shown in Table VIII, the number of critical SNM constraints can be ≤1% of the total number of SNM constraints for the OpenSPARC modules (e.g., for "fgu").

To validate that (26) can be used instead of (25) to compute *PNMV* (i.e., that *PNMV* is unchanged by eliminating all noncritical SNM constraints): we first sample *n* two million times (the number of samples required to estimate $PNMV \le 0.001\%$ for the OpenSPARC modules: details in Fig. 21). Then, we evaluate both $Kn$ and $\widetilde{K}n$ for each sample and verify that (25) is violated if and only if (26) is violated.

*F. Efficiently Solving the MVNCDF Formulation for PNMV*

We make one final adjustment to (31) (in Section III-B2) to further improve the computational efficiency of solving for *PNMV*. In particular, the rows of *C* [in (29) in Section III-B2] can be permuted so that *C* is a block diagonal matrix (i.e., a matrix that has main diagonal blocks that are square matrices and off-diagonal blocks that are zero matrices); for example, starting with the matrix *K* in (47) (in Section IX-D), take the case in which rows 1 and 2 are redundant constraints (e.g., $H_{1,2} < H_{3,4}$ and $H_{2,1} < H_{4,3}$) and remove them from *K* to form $\widetilde{K}$. Then the covariance matrix $C = \widetilde{K}\widetilde{K}^T$ is a block diagonal matrix with diagonal blocks $C_u$ (52).

$$C = \begin{bmatrix} \sigma_1^2 & \sigma_1\sigma_2 & 0 & 0 & \cdots \\ \sigma_2\sigma_1 & \sigma_2^2 & 0 & 0 & \cdots \\ 0 & 0 & \sigma_3^2 & \sigma_3\sigma_4 & \cdots \\ 0 & 0 & \sigma_4\sigma_3 & \sigma_4^2 & \cdots \\ \vdots & \vdots & \vdots & \vdots & \ddots \end{bmatrix} = \begin{bmatrix} C_1 & 0 & \cdots \\ 0 & C_2 & \cdots \\ \vdots & \vdots & \ddots \end{bmatrix}. \quad (52)$$

We now justify why *C* can be block diagonal. Let there be *y* rows of standard cells (e.g., row 1 and row 2 of standard cells are shown in Fig. 8 in Section III-B); each SNM constraint is associated with a single gate pair: $(G^{(dr)}, G^{(ld)})$, where $G^{(dr)}$ is placed in row *u* of standard cells and $G^{(ld)}$ is placed in row *v*. Then the SNM constraints [represented by the rows of $\widetilde{K}$ in (26) in Section III-B2] can be partitioned according to the row of standard cells in which $G^{(ld)}$ is placed, such that the SNM constraints in each partition are independent of the SNM constraints in each other partition. This property holds since:

1) $V_{OH}^{(dr)}$ and $V_{OL}^{(dr)}$ (for $G^{(dr)}$) are independent of the CNT count in the variation-aware SNM model [i.e., $T_{VOH1} = 0$ and $T_{VOL1} = 0$ in (14): Section III-B1]. Thus, the SNM constraints (26) only bound the CNT counts of the CNFETs in the loading logic gate (the driving logic gates affects the tightness of these constraints).

2) The sampling region CNT count variables are partitioned according to the boundaries between the rows of standard cells (e.g., $n_1, \ldots n_6$ belong to row 1 of standard cells in Fig. 8 and $n_7, \ldots n_{12}$ belong to row 2).

The consequence of these properties is as follows. Let $(G_i^{(dr)}, G_i^{(ld)})$ and $(G_j^{(dr)}, G_j^{(ld)})$ be the gate pairs constrained by the *i*th and *j*th SNM constraints, respectively [i.e., the *i*th and *j*th rows of $\widetilde{K}$ in (26) in Section III-B2]. Then the covariance term $C_{i,j}$ is 0 if $G_i^{(ld)}$ and $G_j^{(ld)}$ are placed in different rows of standard cells. Thus, $C = \widetilde{K}\widetilde{K}^T$ can be block diagonal, where the size of each block *u* is equal to the number of SNM constraints in which the loading logic gate is placed in row *u* of standard cells. For example, in (52), $C_1 \in \mathbb{R}^{2 \times 2}$ since there are two SNM constraints that constrain the sampling region CNT counts in row 1 of standard cells.

Since the covariance terms are zero between each block $C_u$ in $C$, then the MVNCDF function in (31) can be separated into a product of smaller MVNCDF functions that are independent: one for each block $C_u$ for each of $y$ rows of standard cells, with corresponding vectors $b_{C_u}$ (such that $C_u$ and $b_{C_u}$ are taken from the same rows of $C$ and $b$, respectively)

$$PNMV = 1 - \prod_{u=1}^{y} \text{MVNCDF}(C_u, b_{C_u}). \quad (53)$$

In our analysis, the size of each $C_u$ is on the order of $\mathbb{R}^{10\times 10}$, corresponding to ~10 gate pairs per row of standard cells (after eliminating noncritical SNM constraints, i.e., following steps 1-3 in Section IX-E). As mentioned in Section III-B2, solving for PNMV in (53) can be efficiently computed, e.g., in less than 10 seconds for the OpenSPARC modules (using a single 2.93 GHz processor core). Table IX summarizes the variables used to compute PNMV. The accuracy of the MVNCDF formulation [both (31) and (53)] is validated in Fig. 21.

## X. GRADIENT DESCENT IMPLEMENTATION

### A. Gradient Calculation

A critical path is any path between a circuit input and a circuit output with propagation delay equal to the maximum path delay. There can be multiple critical paths for a single MC trial (full circuit delay model in Section III-A). Immediately after STA of each MC trial (Fig. 10), we numerically estimate $\nabla T_{95}$ (34) via the following procedure:
1) Record an arbitrary critical path for each MC trial. These paths are used to estimate $\nabla T_{95}$ using a sub-gradient, borrowing from the sub-gradient method for minimization of non-differentiable functions [38] (e.g., "max" in STA).
2) Decrease $p_m$ by an incremental amount (i.e., by $\delta p_m = 10^{-6}$; $10^{-6}$ is <0.1% of all the experimentally demonstrated processing parameter values in Table I), and then recompute the path delay only for the arbitrarily chosen critical path of each MC trial. Build the CDF of these path delays, and extract the delay where the CDF is equal to 95%: this extracted delay value differs from $T_{95}$ by an amount $\delta T_{95}^{(pm)}$. Do the same for $IDC$ and for $p_{\text{Rs}}$ (with $\delta IDC = 10^{-6}$ to compute $\delta T_{95}^{(IDC)}$ and $\delta p_{\text{Rs}} = 10^{-6}$ to compute $\delta T_{95}^{(p\text{Rs})}$).
3) Numerically estimate each element of $\nabla T_{95}$ using (54). This strategy assumes that for each MC trial, the chosen critical path remains a critical path after updating each processing parameter, which in general is only true in the limit as $\delta IDC \to 0$, $\delta p_m \to 0$, $\delta p_{\text{Rs}} \to 0$, and is an approximation for $\delta IDC = 10^{-6}$, $\delta p_m = 10^{-6}$, $\delta p_{\text{Rs}} = 10^{-6}$.

$$\frac{\partial T_{95}}{\partial IDC} = \frac{\delta T_{95}^{(IDC)}}{\delta IDC}, \quad \frac{\partial T_{95}}{\partial p_m} = \frac{\delta T_{95}^{(pm)}}{\delta p_m}, \quad \frac{\partial T_{95}}{\partial p_{\text{Rs}}} = \frac{\delta T_{95}^{(p\text{Rs})}}{\delta p_{\text{Rs}}}. \quad (54)$$

We use a similar methodology to estimate $\nabla PNMV$ (34): for each processing parameter, decrease that processing parameter by an incremental amount and recompute PNMV (to calculate $\delta PNMV^{(IDC)}$, $\delta PNMV^{(pm)}$, and $\delta PNMV^{(p\text{Rs})}$). The elements of $\nabla PNMV$ are estimated using similar equations as (54). We use the same numerical approach to compute $\nabla E_{\text{Tot}}$ (34) [using $E_{\text{Tot}}$ as defined in (13)].

### B. Gradient Descent Step

To update the processing parameters during gradient descent, we first normalize the gradient vector [e.g., $\nabla EDP_{95}$ (35) or $\nabla ENP$ (36)] by its $\ell 1$-norm (i.e., the sum of the absolute values of its elements). This normalizes the magnitude of the step size. We then take a step so that the improvement in each processing parameter is proportional to its corresponding magnitude in the normalized gradient, and so that the total improvement in processing parameters sums to 10%. Small step sizes require more simulations; large step sizes yield coarse granularity in exploration. This strategy (though others may be used) assumes that it is equally difficult to improve each processing parameter by a fixed percentage. For example, if the elements of the normalized gradient have magnitudes 0.70, 0.10, and 0.20, then we reduce $IDC$, $p_m$, and $p_{\text{Rs}}$ by 7%, 1%, and 2% (versus their current values), respectively.

### C. Avoiding Convergence to Local Optima

Given that our optimization methodology is based on gradient descent, we employ two strategies to avoid convergence to local optima (since the objective is not necessarily a convex function):
1) *Initialize Gradient Descent from Multiple Design Points on the Initial Energy-Delay Tradeoff Curve*: each instance of gradient descent typically leads to a unique design point. Even if all instances of gradient descent converge to the same local optimum, we never choose a worse design point by starting another instance of gradient descent.
2) *Never Increment a Processing Parameter Away from its Ideal Value* (Table I): The only case in which all CNT count variations are zero, by definition, is the nominal case (in which all processing parameters have their ideal values). This is consequently the global optimum in terms of minimizing the effect of variations. Any case in which the gradient vector points toward incrementing the value of a processing parameter away from its ideal value is indicative of local optima; we choose to not update that parameter in these cases.